\documentclass[twocolumn]{aastex631}
\usepackage{listings}
\usepackage{amsmath}

\newcommand{\planetarium}{Plan\'etarium de Montr\'eal, Espace pour la Vie, 4801 av. Pierre-de Coubertin, Montr\'eal, Qu\'ebec, Canada}
\newcommand{\irex}{Trottier Institute for Research on Exoplanets, Universit\'e de Montr\'eal, D\'epartement de Physique, C.P.~6128 Succ. Centre-ville, Montr\'eal, QC H3C~3J7, Canada}

\begin{document}

\title{21,864 Unresolved, Low-mass Binaries Identified via their Overluminosity in \textit{Gaia} DR3 and a Catalog of 347,440 Systems within 100 pc of the Sun
}

\author[0000-0003-0179-9662]{Zachary Way}
\affiliation{Department of Physics and Astronomy, Georgia State University, Atlanta, GA 30303, USA}

\author[0000-0002-2437-2947]{S\'ebastien L\'epine}
\affiliation{Department of Physics and Astronomy, Georgia State University, Atlanta, GA 30303, USA}

\author[0000-0002-2592-9612]{Jonathan Gagn\'e}
\affiliation{\planetarium}
\affiliation{\irex}

\author[0000-0003-3410-5794]{Ilija Medan}
\affiliation{Department of Physics and Astronomy,
	Vanderbilt University,
	Nashville, TN 37235, USA}

\begin{abstract}

The fundamental parameters of a low-mass star can potentially be determined from its photometry and astrometry. This is complicated by the fact that 10-20 percent of low-mass stars are predicted to be equal-mass binaries. These unresolved systems appear more luminous compared to single stars with the same fundamental parameters. We present a method to differentiate binary stars from single-star main sequence K and M dwarfs using their \textit{Gaia} DR3 XP spectra. We assemble a training set of stars which have pristine astrometry and photometry, are located within 100pc of the Sun, and exclude stars with \textit{Gaia} DR3 flags suggesting they may be unequal mass systems, thereby leaving stars that are predominantly either single- or equal-mass binaries. We then iteratively train Random Forest Regression (RFR) models to predict absolute magnitude and color given the RP spectral coefficients of a star. After each model, we remove the stars that have absolute magnitudes significantly brighter than their predicted values. This method converges on a model trained only on single stars. We then use this model to identify the ``overluminous'' K and M stars in \textit{Gaia} DR3 within 100 parsecs, with some quality cuts. We find that $\sim13\%$ of the sample is significantly overluminous and assume these to be unresolved binaries. 
We aggregate several multiplicity surveys across different projected separations and incorporate our overluminous binaries to create a general \textit{Catalog of Systems} within 100 pc. We use this \textit{Catalog} to provide lower limits on the multiplicity fraction for stars between $0.1$ and $0.7~M_{\odot}$. 

\end{abstract}


\section{Introduction} \label{sec:intro}

\begin{figure*}
    \centering
    \includegraphics[width=\textwidth]{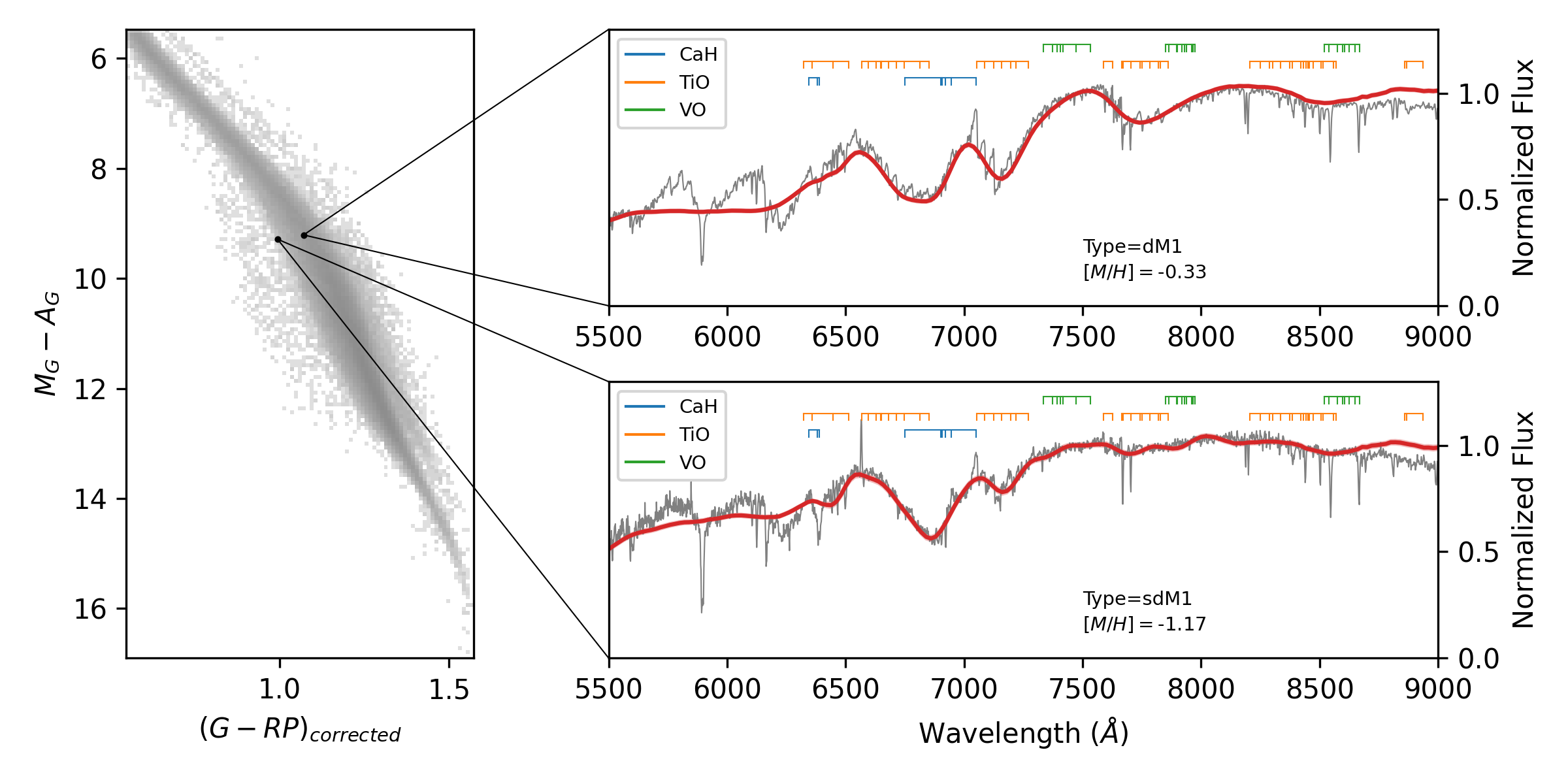}
    \caption{The strong relationship between the optical photometry and spectra for low-mass stars. Left panel: the dereddened, \textit{Gaia} color-magnitude diagram for our stellar sample defined in \S\ref{sec:data}. Two examples, a metal-rich dwarf (PM J08202+0532) and a metal-poor subdwarf (G 192-52), are pulled from this distribution. Right panels: the low-resolution, XP spectrum (red) and a medium-resolution, LAMOST DR9 (gray) spectra for these examples. Their location on the CMD is shown and arrows point to their spectra on the right panels. Labels above the spectra identify the strongest molecular features in the SED showing that there are drastic changes for different metallicities. This change is strong enough to effect the photometry. }
    \label{fig:hr_and_spectra}
\end{figure*}

Binary systems consisting of low-mass (red) dwarfs are the most numerous multiple stars in our Milky Way but have historically been difficult to identify. Due to their inherent dimness and cool, red atmospheres, the most complete multiplicity surveys of low-mass stars only stretch out to a few tens of parsecs \citep{recons_volume_complete, tokovinin_multiple_stars, pokemon_survey}. Other multiplicity surveys use specific methods to evaluate binarity, each with its own strengths and limitations. The first method is to identify stars with common proper motions \citep{superwide, elbadry_WB} and visual binaries \citep{medan_visual_binaries, WDS}. These methods are robust at wide separations, but limited by the survey's angular resolution, and is therefore best at probing long period systems \citep{binary_renaissance}. For the \textit{Gaia} \citep{Gaia} imaging survey, the angular resolution limits the identification of binaries to $\gtrsim 0.5$ arcseconds. Perturbations in the astrometric solutions can however indicate the presence of an unresolved companion to separations below the nominal Gaia resolution limit \citep{belokurov_astrometry,carmenes_multiplicity, binaries_GU}. 

At even closer separations, binary systems can be systematically identified using high-resolution imaging techniques such as speckle imaging \citep{DSSI, QWSSI, medan_speckle, pokemons_speckle, nessi_binary}, adaptive optics \citep{roboAO_catalog, roboAO_lowmass}, or long-baseline interferometry \citep{mircx_chara, torres_chara_binary, zhao_chara_resolved_binary, gallene_vlti_orbits, vlti_brown_dwarf}. These methods can reach separations $<0.5$ milliarcseconds, but are limited to bright targets. Thus, systematic studies with these methods have been limited. 

Unresolved binary systems may be identified as eclipsing systems \citep{asassn_variable_stars, ztf_variable_stars, tess_ebs} or by their variation in radial velocity \citep{kounkel_SB2, delta_rv_max}. These methods probe closely separated binaries ($\lesssim 100~AU$), but are subject to projection effects, where systems cannot be identified when viewed from the top of the orbital plane.

Another, more general method is to identify binaries based on ``overluminosity'', i.e. when a star appears more luminous than other objects with similar fundamental parameters, pointing to the presence of an unresolved companion \citep{binaries_on_HRD, lobster}.  Traditionally, these systems have been identified in observations of star clusters, where a second ``binary main sequence'' is typically found to hover above the theoretical single-star main sequence \citep{mesa_brandner}. Since all stars in a cluster are born at the same time, a single star and an equal-mass binary system will have the same color but differing luminosities. This method has been used to identify low-mass binaries in open clusters such as the Hyades \citep{hyades_bonafide}. However, cluster fitting heavily relies on the prior supposition that all stars in the cluster lie along a well-defined isochrone, which can be predicted from models but typically depends on metallicity, age, and interstellar reddening. Here, a question arises: ``can this method be applied to field populations of low-mass stars?" Historically, the answer was ``no'' because we have not had reliable measurements of luminosity for non-cluster stars, but the availability of reliable parallaxes from \textit{Gaia} allows us to calculate absolute magnitudes for most relatively nearby field stars.

In order to determine overluminosity for stars in field populations, it is first necessary to know what a star's expected luminosity should be, which normally would require prior knowledge of the star's mass, metallicity, and age. 
Previous studies have used model predictions of photometry or stellar labels measured from spectroscopic datasets. \cite{wallace_binaries} utilizes the former method by predicting the SED and photometry of sources along a model isochrone. For each star, a mass ratio (q) is predicted from the estimated composite photometry predicted by the models. \cite{liu_binary_stars} uses spectroscopy from LAMOST DR10 rather than photometry to identify binarity by fitting a binary fraction and a mass-ratio power law to the data in bins of metallicity ([Fe/H]). In a similar study,  \cite{lamost_overluminosity} train a convolutional neural network (CNN) to predict the probability that a system is a binary within the dataset defined in \cite{liu_binary_stars}; binary stars were labeled as overluminous by binning the sources by metallicity and selecting those that lie above the main sequence. Most recently, \cite{li_binaries} identified binaries using the \textit{Gaia} XP spectra. This work builds a forward model of a star's XP spectrum given its absolute magnitude, $BP-RP$ color, and metallicity. Because the XP spectra have exquisite absolute flux calibration, it is possible to fit each source's absolute flux and become very sensitive to the presence of faint companions \citep{XP_carrasco, XP_montegriffo}.

These methods have found success for the higher mass systems ($\gtrsim0.33 M_{\odot}$) but struggle for the red dwarfs. Unfortunately, systematic biases remain for low-mass stars because their models are still unable to predict accurate spectroscopy and photometry \citep{Torres_mdwarf, morrel_mdwarf, mesa_brandner, mann_mdwarf}. Low-mass stars are difficult to model as they often lie in the uncomfortable regime of noisy data and difficult radiative transfer modeling due to their cool temperatures \citep{Osterbrock1953, BT-Settl}. Unlike higher-mass systems, molecular bands dominate their SED and, as a result, these systems do not have reliable stellar parameters \citep{Bergemann2017}. Previous methods for predicting overluminosity, or any broad analysis of low-mass, stellar parameters, fail at absolute magnitudes ($M_G$) greater than $\sim10$ \citep{andrae_xp, medan_xmatch}.

However, while difficult to model, the spectra of cool dwarfs have been shown to be incredibly dynamic for different fundamental parameters, where a small change in $T_{eff}$ and [M/H] results in large variations in molecular band strengths and ratios \citep{lepine_07, lepine_13, hejazi_metallicity}. These changes also significantly alter the luminosity of an object. For example \cite{kesseli_metal_poor}, showed that a $\sim 1~dex$ reduction in metallicity for a star of the same stellar mass can lead to a $\sim$ 20\% reduction in radius. This leads to a drastic change in surface temperature for these systems, and the bluing of the overall SED for more metal-poor objects causes the low-mass main sequence of the field star population to be much broader than for higher masses, due to general metallicity variations among field stars. These effects can be seen in Figure \ref{fig:hr_and_spectra} where the \textit{Gaia} XP and LAMOST spectra are shown for two stars with a similar absolute G-band magnitude. The molecular species dominant in this regime are displayed in each spectrum: titanium oxide (TiO), calcium hydride (CaH), and vanadium oxide (VO). As metallicity decreases, there is a drastic reduction in the diatomic, metal-oxide features compared to the hydrides as they scale quadratically with the overall metallicity (with two metal atoms per molecule) rather than linearly. This reduction is so strong that it can even be seen in very low resolution. The medium resolution spectra ($R\sim2,000$) are from LAMOST DR10\footnote{https://www.lamost.org/dr10} \citep{lamost, lamost_dr} while the low resolution are the corresponding \textit{Gaia} XP spectra ($R \sim 50 - 100$). Strong changes in molecular absorption are reflected in both surveys, implying that the XP spectra contain meaningful information regarding the metallicity and temperature of these stars. Thus, while the measured parameters from model fitting are not trustworthy, there remains a strong relation between a star's position on the CMD and its measured spectrum.

In this paper, we propose to expand this approach and leverage the very large set of K/M dwarfs with XP spectra and accurate luminosities measured from \textit{Gaia}, circumventing the need for predetermined fundamental parameters when predicting the luminosity of a K/M dwarf. We use simple Decision Tree Regression methods to predict the CMD location for stars directly from their spectra. Then, we use this prediction to search for overluminosity and identify candidate binary systems in the field population. In \S\ref{sec:data} we describe the \textit{Gaia} XP spectra and define our sample of low-mass stars. In \S\ref{sec:method} we describe how we identify overluminous binaries using a decision tree regression model. We validate our sample of binaries in \S\ref{sec:validation}, showing that we recover a majority of binaries from known catalogs. In \S\ref{sec:catalog}, we curate a catalog of multiple systems given our predictions and membership in other multiplicity catalogs and show our multiplicity rates in \S\ref{sec:results}. Finally, in \S\ref{sec:summary} we summarize our work and discuss our results in the context of other multiplicity surveys.

\begin{figure*}
    \centering
    \includegraphics[width=\textwidth]{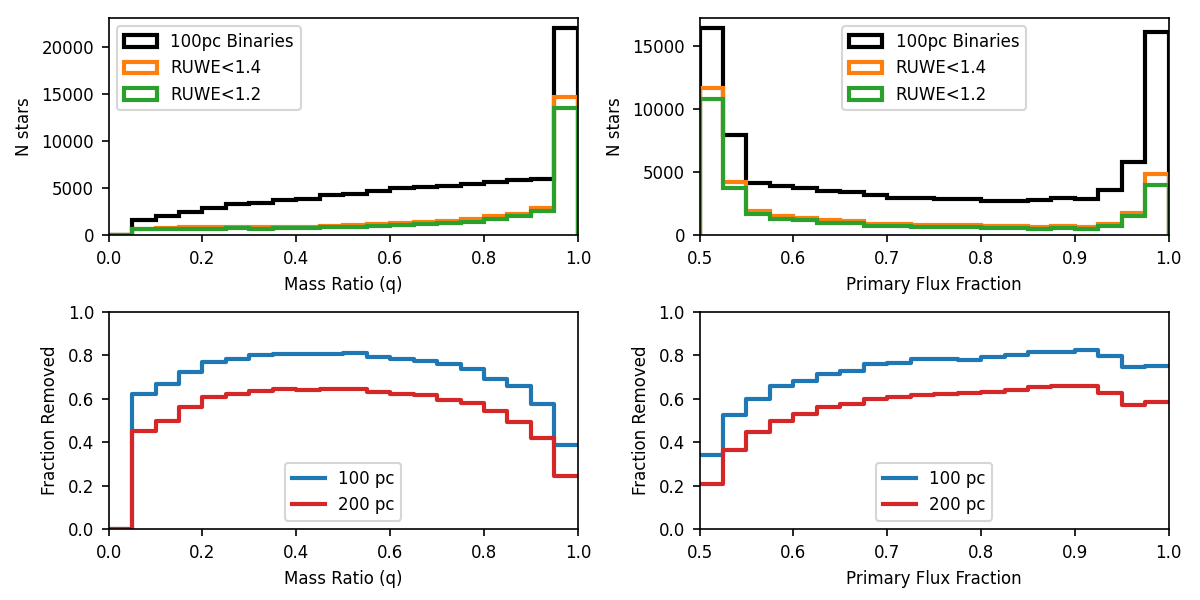}
    \caption{The results of different RUWE cuts for low-mass stars from a simulated set of 200,000 binaries. The predicted RUWE parameters were generated using the \textsc{GaiaUnlimited} \citep{empirical_GU, subsample_GU} package. A linear interpolation was used to relate mass and absolute magnitude from \cite{the_bible}. Left panels: the top panel displays the mass ratio distribution of our simulation, as found by \cite{elbadry_twin}. The resulting mass ration distribution from two RUWE limits, $1.4$ in orange and $1.2$ in green, are also shown. The bottom panel shows the fraction of stars removed at a given mass ratio for different distance limits using the constraint RUWE$<1.2$ which we adopt in \S\ref{sec:data}. Right panels: the distribution (top panel) and fraction of stars removed (bottom panel) for a given primary flux fraction for the full sample and the stars below the RUWE thresholds. The primary flux fraction is defined as the amount of flux emitted from the system which comes from the brigther star. The RUWE cuts are less effective at removing the equal-mass systems ($\sim0.5$) and effectively single stars ($\sim1.0$).}
    \label{fig:RUWE}
\end{figure*}

\section{Data and Sample Selection} \label{sec:data}

We utilize the \textit{Gaia} XP spectra to make predictions about stars' overluminosity. These data were collected by the European Space Agency's \textit{Gaia} space telescope and are low-resolution ($R\sim50-100$), prism spectra \citep{XP_carrasco, XP_de_angeli}. There are two instruments, the BP and RP spectrographs, which cover a range of $330-680~nm$ and $640-1050~nm$, respectively, and are collectively referred to as the XP spectra. 
The \textit{Gaia} team represents the mean spectrum of a star not as flux measured at a specific wavelength but as a decomposition into a Hermitian basis, with 55 coefficients representing the spectrum in each band.

Our method relies on precise measurements of each star's absolute magnitude ($M_G$), $G-RP$ color, and $RP$ spectrum. Thus, stringent requirements were made to define our sample of stars. We select K- and M-type dwarfs by requiring $G-RP>0.56$ and absolute magnitude $M_G>5.553$, consistent with values from \cite{the_bible}. We also adopt a color-dependent magnitude limit $M_G<10 (G-RP)+5$ to remove white dwarfs from the sample. To avoid contamination due to reddening and stars with poor distances, we select stars within 100 parsecs ($\varpi>10$) and with low parallax error ($\varpi/\delta\varpi > 10$). Some XP spectra have been contaminated by nearby sources. To remove those, we also require that the RP spectrum is blended or contaminated less than 10\% of the time. \cite{XP_de_angeli} describe blending as ``when two or more sources fall within the observed window'' and contamination as when the spectrum contains ``flux belonging to a source that is located outside the window.''

\begin{figure*}
    \centering
    \includegraphics[width=\textwidth]{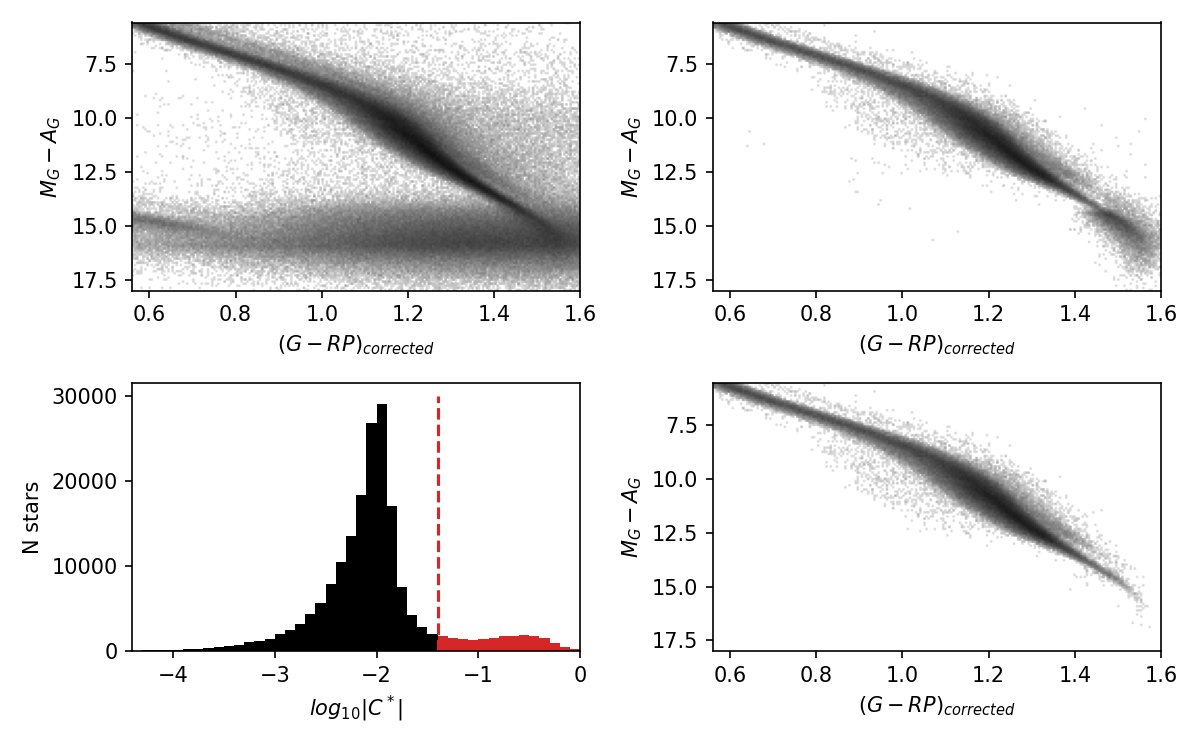}
    \caption{The winnowing from all stars within 100 pc to the Regression Sample in four parts. Top left: the CMD of all low-mass stars in \textit{Gaia} with a parallax $\varpi>10~mas$. Top right:  the resulting sample given our \textsc{ADQL} query described in \S\ref{sec:data}. It can be seen that there are still issues with the photometry, especially in the lowest-mass stars. Bottom left: the distribution of the corrected \textsc{phot\_bp\_rp\_excess\_factor}, which we represent as $C^*$ \citep{riello_photometry}. The red line at $log_{10}|C^*| = -1.4$ represents our cut for stars with poor photometry. Bottom right: the \textit{Gaia} CMD with the high $C^*$ sources removed, representing the final Regression Sample of 127,843 stars.}
    \label{fig:sample}
\end{figure*}

Another common parameter used to remove stars with possibly dubious parallaxes is \textit{Gaia}'s Renormalized Unit Weight Error (RUWE), which is a renormalized version of the $\chi^2$ parameter \citep{lindegren_RUWE}. RUWE serves both as a quality cut on the astrometric solution of a given system and a signal that a system's photocenter does not follow a single star's trajectory in the sky. Stars with high RUWE have been interpreted as binary stars, but, critically, not all binary stars are expected to have high RUWE values \citep{lindegren_binaries}. Stars with RUWE close to unity are not necessarily guaranteed to be single, and binaries of near equal-mass are not expected to show significant wobble in their photocenter from orbital motion \citep{belokurov_ruwe, empirical_GU}. Furthermore, it has been found that close binaries with low-mass stars have a preponderance of equal-mass systems \citep{elbadry_twin}. This implies that a large fraction of binaries are not identified by a selection based on RUWE, especially for the nearby, low-mass stars. 

\begin{table}[]
    \centering
    \begin{tabular}{l|r}
         Parameter & Distribution \\
         \hline
         Mass Ratio (q) & Mass Ratio Dist.  \\
          & ($0.1 < M/M_{\odot} < 0.4$)\textsuperscript{1} \\
         Primary Mass [$M_{\odot}$] & $U(0.1,0.4)$\\
         Distance [pc] & $U(5,200)$\\
         log ( Period [yr] )  & $U(-2.56, 2)$\\
         Eccentricity & $U(0,1)$\\
         Initial Phase [rad]  & $U(0,2 \pi)$\\
         Azimuthal view angle ($\phi$) [rad] & $U(0,2 \pi)$\\
         cos ( Polar view angle ($\theta$) [rad] ) & $U(-1,1)$\\
         Planar projection angle ($\omega$) [rad]  & $U(0,2 \pi)$\\
         Right Ascension [deg] & $U(0,360)$\\
         cos (Declination [deg] ) & $U(-1,1)$\\ 
    
    \end{tabular}
    \tablenotemark{\textsuperscript{1}distribution taken from \cite{elbadry_twin}}
    \caption{Parameters of the distributions from which the synthetic binaries were generated.}
    \label{tab:synthetic_dist}
\end{table}

To demonstrate this principle, we use the \textsc{GaiaUnlimited}\footnote{https://github.com/gaia-unlimited/gaiaunlimited} \citep{subsample_GU, empirical_GU} package to test which binaries would be removed by a RUWE cut. \textsc{GaiaUnlimited} predicts measured parameters in \textit{Gaia} DR3 given a simulated source's sky location and brightness \citep{empirical_GU}. The software was extended to allow a user to model the predicted RUWE of a given binary system. \cite{binaries_GU} demonstrated this feature and predicted which fraction of binaries will have an enlarged RUWE given a uniform distribution of mass ratios. To show the effect that a RUWE cut will have on our sample of low-mass stars we use their same analysis but adopt physically motivated mass ratio distribution (i.e. one with a larger number of equal-mass systems) from \cite{elbadry_twin}. We generate 200,000 binaries with random parameters following the distribution described in Table \ref{tab:synthetic_dist}. Then, we approximate the mass-absolute magnitude relation as a monotonic interpolation according to \cite{the_bible}. Finally, we find the predicted RUWE for this sample using \textsc{GaiaUnlimited}. The results of this model are shown in Figure \ref{fig:RUWE} with the effects of a 1.4 and 1.2 RUWE rejection cut.
It can be seen that, while most of the unequal-mass systems are removed after these RUWE cuts, more than half of the equal-mass systems remain. For these systems then, a RUWE rejection cut is most effective at removing systems that have a companion which contributes a significant amount of light to the system. However, if the companion is too similar in brightness to the primary, then a RUWE rejection cut becomes inefficient again. Unfortunately, this equal-mass regime is the most abundant mass ratio of systems for this low-mass range ($M\in[0.1,0.4]~M_{\odot}$). As \cite{penoyre_ruwe} state in their abstract, ``Most binaries are undetected." This is illustrated in the right-hand panels in Figure \ref{fig:RUWE}. Systems with a primary flux fraction near 1.0 are effectively like single stars, where the unseen companion is not visible in the SED.  Systems with a primary flux fraction of 0.5, on the other hand, are the equal-mass binaries. Both types tend to remain in significant numbers after the RUWE cuts. 

In this study, we limit our sample to those stars with $RUWE<1.2$ with the expectation that these will either be single stars, pairs with a flux-dominant primary looking like a single star spectroscopically and photometrically, or equal-mass binaries that have twice the flux of their single-star analogs. This is a fundamental assumption for our method in \S\ref{sec:method}.

The ADQL query which generates our resulting sample of nearby ($d<100$pc) low-mass main-sequence stars, with uncontaminated \textit{Gaia} XP spectra, with high quality parallax and colors, and with low RUWE values can be found in Appendix \ref{appendix:regression_sample} and results in 138,600 stars.
Their color-magnitude diagram (CMD) is compared to that of the full \textit{Gaia} 100 pc sample in Figure \ref{fig:sample}.
The main sequence from our query is much better defined than that of all stars with $\varpi>10~mas$, which shows that most stars with problematic astrometry/photometry have been excluded. However, a number of stars with inconsistent data remain. First, there is a large amount of scatter in objects with absolute magnitude greater than $\sim13$.  Second, there are several stars that are far too blue or red for their absolute magnitude. These objects can be seen to the far left or right of the main sequence. We attribute these objects with odd CMD locations to a problem with the RP photometry in \textit{Gaia} DR3.
To address this issue, we use the photometric BP/RP color excess for each source, defined as $C=(I_{BP} + I_{RP})/I_G$ where $I_i$ is the total flux in a certain passband. This value should be nearly constant because the BP and RP passband responses cover the G-band response. However, $C$ has a dependence on color for low-mass stars. \cite{riello_photometry} provide the following correction to the color excess $C^* = C + \sum a_i (BP-RP)^i$, where $a_i$ is given in Table 2 of their article. The distribution of $log_{10}|C^*|$ is shown in Figure \ref{fig:sample}. There are two groups of stars, those with low ($ log_{10}|C^*|\sim-2.4$) and high ($ log_{10}|C^*|\sim-0.5$) color excess. The latter group is likely due to noisy photometric measurements or because the source has blended RP photometry, where the G-band measurement is resolved but the larger BP/RP aperture contains flux from a nearby source. These partially resolved targets can be removed with the selection $log_{10}|C^*|>-1.4$ leaving the 127,843 stars shown to the lower right of Figure \ref{fig:sample}. We dub this group of stars the Regression Sample.

Finally, because we want to find the relationship between a star's XP spectrum and its intrinsic absolute magnitude and color, we correct $M_G$ and $G-RP$ for interstellar reddening. We use the \textsc{dustmaps} \citep{dustmaps} package to find the integrated extinction based on the 3D dust map of \cite{edenhofer_map}.  We then transform the extinction into the \textit{Gaia} passbands using the following relations based on Table 3 of \cite{extinction_relation} and \cite{kim_metallicities}
$$E(BP-RP) = 1.320~E(B-V)$$
$$A_G = 1.890~E(BP-RP)$$
$$E(G-RP) = 0.461~E(BP-RP)$$
In all plots, we show the dereddened photometry where $(G-RP)_{corrected} = (G-RP) - E(G-RP)$. The \cite{edenhofer_map} extinction map is only constrained between 69 pc and 1.25 kpc, so we set $E(BP-RP)=0$ for sources that are closer than the inner boundary of the dust map. This is a simplistic approach to dereddening and the conversion from $E(B-V)$ to $E(G-RP)$ should depend on spectral type \citep{mu_tau_gagne}. However, the extinction for these nearby sources should be negligible, and, in any case, 99\% of our sample has an estimated reddening of less than $0.03$ mag and the maximum reddening is $0.14$ mag.

\section{Color-luminosity prediction with Random Forest Regression} \label{sec:method}
\subsection{Prior assumptions for the Regression Sample}\label{sub:100pc}

\begin{figure*}
    \centering
    \includegraphics[width=\textwidth]{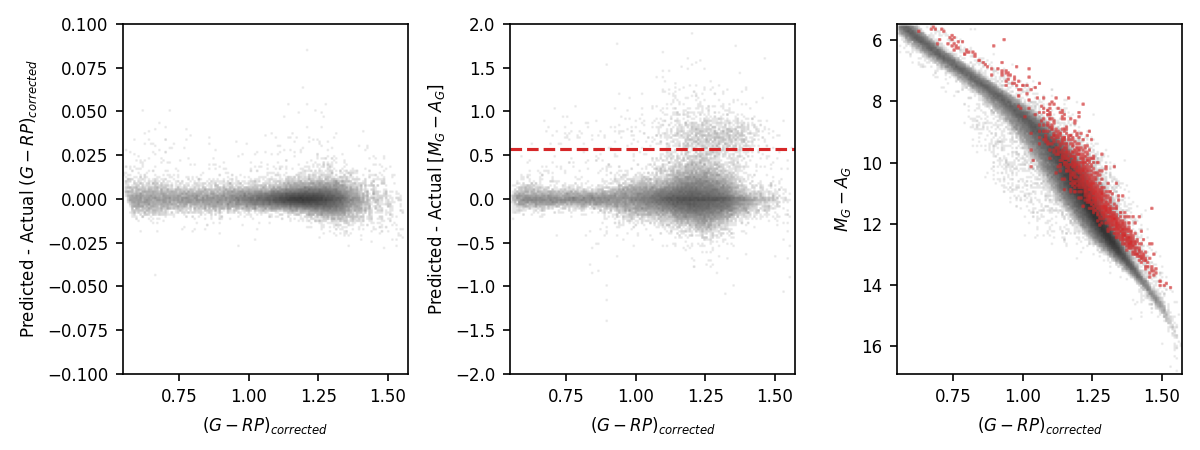}
    \caption{A comparison between the predictions from our first \textsc{HistGradientBoostingRegressor} model and the measured values from \textit{Gaia}. Left panel: the precision of the predicted color compared to the measured values. The low scatter shows that we recover this feature well. Middle panel: the precision of the absolute magnitude prediction. Unlike the color, there is a cloud of stars hovering above the central distribution. These systems have a lower absolute magnitude and are therefore brighter than their prediction: the overluminous binary systems. We identify these systems with the dashed red line which lies at 3 standard deviations above zero. Right panel: the overluminous sources are highlighted in red on a \textit{Gaia} CMD. }
    \label{fig:first_model}
\end{figure*}

After the selection made in \S \ref{sec:data}, the Regression Sample is composed primarily of single and equal-mass binary stars which have clean XP spectra, absolute magnitudes, and colors. In order to separate the binaries from the single population, we utilize three assumptions: (1) stars close to each other on the CMD should have similar XP spectra, (2) the SEDs of unresolved binaries and single-stars of the same physical parameters are degenerate, except that they emit twice the flux, and (3) most low-mass stars in the field are single.

The first assumption should be valid for single stars in our Regression Sample. It is possible for reddening to shift a star's location on the CMD, making it appear redder and fainter. However, we account for reddening in \S\ref{sec:data} and most of our sources have negligible reddening. Other internal effects may change a star's location on the CMD and break the relation to its fundamental parameters. For example, fast rotation may induce spotting on the surface of the star (causing a redder color) and young stars are more luminous than their older counterparts. Due to their long lifetimes, we expect the field population of M dwarfs we expect most stars to be older than 1 billion years and for their rotation to be very slow \citep{lu_rotation}. 

The second assumption should be valid for all equal-mass binaries, but may not hold true for unequal mass systems, where you will measure the summed flux of two stars with different effective temperatures. It will however hold for the majority of stars in the Regression Sample because many unequal binaries would have been removed with our RUWE rejection cut in \S\ref{sec:data}. 

Given the first two assumptions, a model can be trained which predicts the absolute magnitude and color of a star given its XP spectrum. And, if the model is trained only on the single stars in the sample, its predicted absolute magnitude will be incorrect for the binary systems because they are twice as luminous. By measuring the binaries' overluminosity they can be identified in the field population. Unfortunately, there is no robust catalog of binary stars reaching further than $20$ pc so we do not have labels with which we can identify true-positive binaries. We cannot utilize traditional labeled regression methods and instead must use another approach to distinguish these stars. 

The third assumption is supported by volume complete surveys of the Solar neighborhood and is an aspect that we can leverage to identify the binaries \citep{recons_volume_complete}. A model that is trained on a subset of the Regression Sample will indeed have binaries, but given the single-binary degeneracy, the predicted absolute magnitude will be an average of the single and binary star values. Then we can compare the measured values to the predicted values and see that the binaries are systematically too luminous. By removing these overluminous stars and training a new model on the stars with ``correct" predictions, we can iteratively clean the binary stars from our sample. Before we can iterate, though, we must first define our regression problem.

\begin{figure*}
    \centering
    \includegraphics[width=\textwidth]{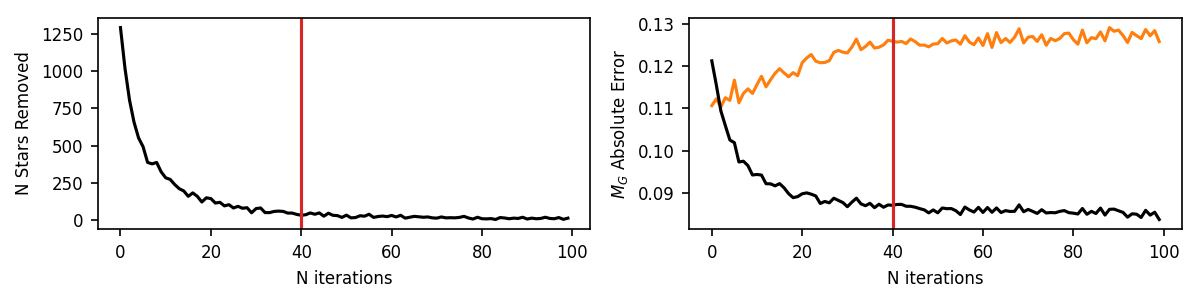}
    \caption{Progress of our iterative method shown in two aspects. Left panel: number of stars flagged as ``overluminous" and removed at each iteration. Right panel: absolute error in the predicted absolute magnitude after each iteration. The error for each iteration's test sample is shown in black and the error for the full sample (including the overluminous systems) is shown in orange. We can see that there are progressive gains but with diminishing return. We note the elbow in the total sample's error, where we identify that each iteration is no longer removing binaries. We mark the 40th iteration as the breaking point by eye and use this as our terminal model.}
    \label{fig:iterations}
\end{figure*}

\subsection{Regression Model Definition}\label{sub:reg_model}

We define a model which can predict a star's position on the CMD (absolute magnitude and color) given its XP spectrum. We elect only to use the RP spectra because while \cite{poor_old_heart} found that including the BP coefficients marginally helps their ability to predict metallicity for red giants, most of the red dwarf sources in our Regression Sample have very low signal-to-noise BP spectra.
We then choose to normalize the 55 RP coefficients (which have units $e^-/s$)
by the total integrated flux of the RP spectrum such that
\begin{equation}
    \boldsymbol{c_i^*} = \frac{\boldsymbol{c_i}}{\int \left[ \sum_{k=0}^{54} c_{i, k} \cdot F_k(\lambda) \right] d\lambda}
\end{equation}
where $\boldsymbol{c_i}$ is the vector of RP Hermitian coefficients for the i-th star in the Regression Sample ($\boldsymbol{c_i}=[c_{i,0}, c_{i,1}, ..., c_{i,54}]$), $F$ is the design matrix of Hermitian basis functions and $\boldsymbol{c_i^*}$ is the vector of normalized coefficients which serve as our input data for the model. These normalized coefficients are now stripped of information related to the total flux of the object, describing only the {\em morphology} of the spectrum. Next, we normalize our target data by subtracting the mean and scaling to the unit variance.
\begin{equation}
    \boldsymbol{y_i^*} = \frac{\boldsymbol{y_i} - \boldsymbol{\bar{y}}}{\boldsymbol{\sigma_{y}}}
\end{equation}
Here, $\boldsymbol{y_i}$ is the vector of target parameters for the i-th star defined as 
\begin{equation}
    \boldsymbol{y_i}=
    \begin{pmatrix} 
    [M_G - A_G]_i \\ 
    [(G-RP)_{corrected}]_i 
    \end{pmatrix}
\end{equation}
$\boldsymbol{\bar{y}}$ is the average value for each value over the full Regression Sample, $\boldsymbol{\sigma_{y}}$ is the standard deviation, and $\boldsymbol{y_i^*}$ are the rescaled final target parameters. 

We tested two decision tree regressors in \textsc{scikit-learn v1.6.1} \citep{scikit-learn}, \textsc{HistGradientBoostingRegressor} and \textsc{RandomForestRegressor}, using their \textsc{RandomizedSearchCV} method to test a grid of hyperparameters for each model. We vary \textsc{max\_features}, \textsc{max\_leaf\_nodes}, and \textsc{min\_samples\_leaf} for the \textsc{RandomForestRegressor}. We vary \textsc{max\_iter}, \textsc{max\_depth}, and \textsc{learning\_rate} for the \textsc{HistGradientBoostingRegressor}. 
We find that these two decision tree regressors perform similarly well, but the \textsc{HistGradientBoostingRegressor} produces a slightly lower total absolute error and trains much faster. Similar studies have also preferred this model when performing regression with the XP spectra \citep{andrae_xp}. We settle on using two separate \textsc{HistGradientBoostingRegressor} models, one that predicts the absolute magnitude $M_G-A_G$ and one that predicts the color $(G-RP)_{corrected}$. Each model has a maximum tree depth of 10, a maximum iteration limit of 5000, and a learning rate $\sim$0.0163. We train two models simultaneously because models predicting more than one variable are not supported in the \textsc{HistGradientBoostingRegressor} method and it is qualitatively similar to a decision tree which predicts both values at once.

\subsection{Training the model}\label{sub:train_model}

We split the Regression Sample into a training set and a testing set at a $2:1$ ratio and do an initial pass at training our regressor. The predictions of this model for stars on the testing set are compared to the observed values in Figure \ref{fig:first_model}. We are recovering the color very well from the RP coefficients. But, for the absolute magnitude, there is a secondary cloud of stars hovering above the main distribution, with predicted absolute magnitudes significantly lower than the measured average values for a given color. These stars are the overluminous systems. We label them as overluminous in the training set by finding stars which lie 3 standard deviations above zero in in the Predicted - Actual [$M_G - A_G$] distribution. While these systems are marked in the testing set, there still remain binaries in the training set.

We remove these overluminous systems from the overall sample and randomly redistribute our remaining stars into new testing and training sets, keeping the same $2:1$ ratio. We then repeat the procedure and train the regressor on the updated sets, and we remove additional overluminous systems, i.e., stars lying 3 standard deviations above zero. It is important to note that we only label stars as overluminous when they are used in the testing set, to avoid biasing the convergence of the model. Our iterative method is defined by the following:
\begin{enumerate}
    \item Split the sample into a training/testing set with a ratio of 2:1
    \item Train models to predict the color and absolute magnitude
    \item Predict the absolute magnitude of the testing set and compare with \textit{Gaia}'s measurement
    \item Remove stars which are more than 3 standard deviations brighter than their prediction
\end{enumerate}
This process was repeated for 100 iterations. In Figure \ref{fig:iterations} we show the number of stars removed after each iteration. We also show the mean absolute error for the entire stellar sample compared to the error from each consecutive test set. Initially, many stars are removed as overluminous but near the end we are removing fewer than 50 stars per iteration. 
This is mirrored in the absolute error where there is an initial steep convergence followed by diminishing gains in error. The opposite is seen for the entire sample, which still contains the binaries that are progressively removed from the training set. Visual inspection reveals that the later models are removing potentially important outlier stars as ``overluminous" (notably some of the obvious metal-poor subdwarfs) and an earlier iteration needs to be set as the terminal model to avoid removing critical objects. We define this terminal model using the break in absolute error of the entire sample on the 40th iteration. Before this break, the absolute error degrades because we are removing unresolved binaries, which have a significant difference in absolute magnitude. But after iteration 40, we see minimal degradation, indicating that we are only removing stars that are distant from the mean distribution (e.g. the subdwarfs).

\begin{figure*}
    \centering
    \includegraphics[width=\textwidth]{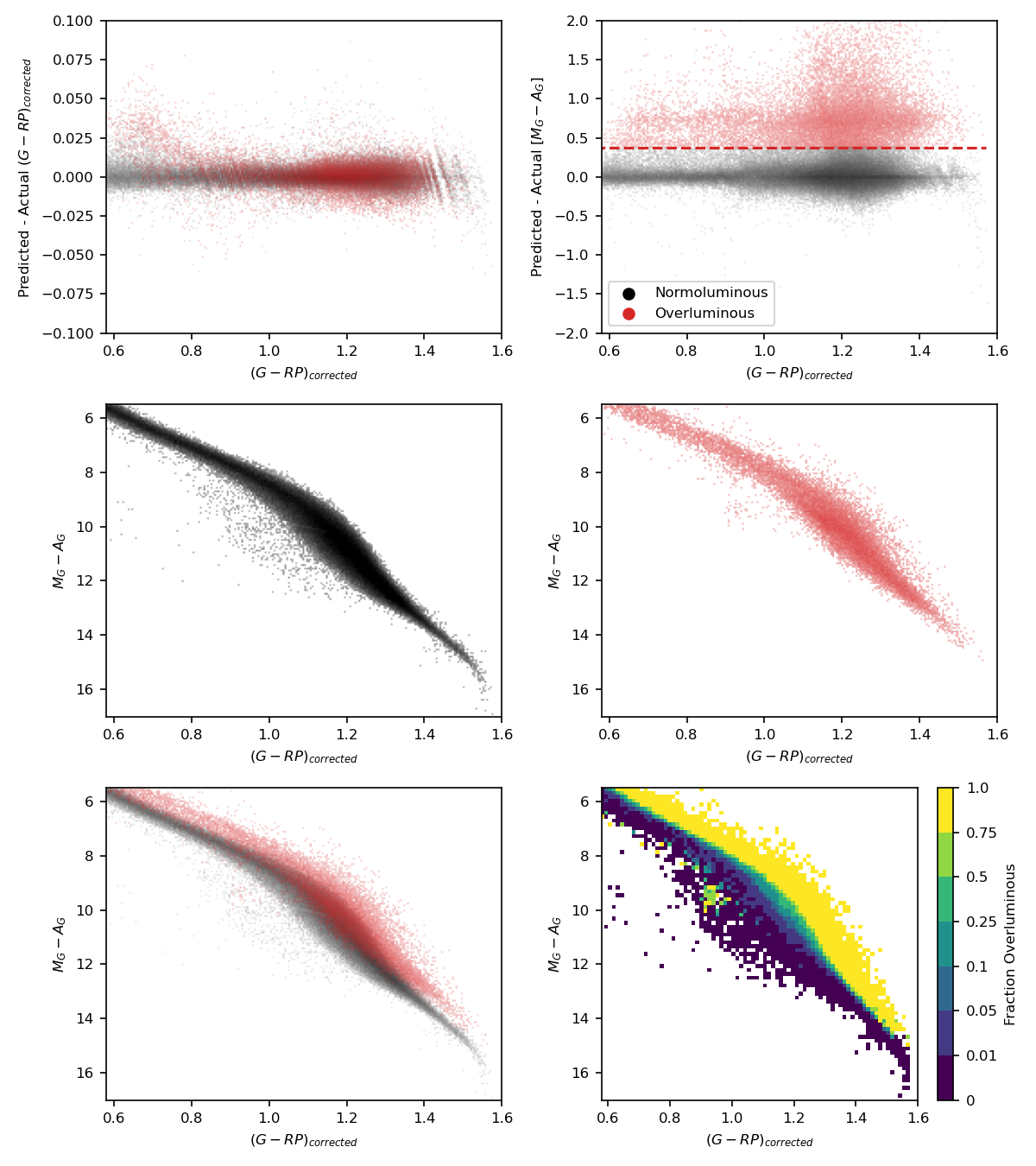}
    \caption{Predictions from our terminal model (iteration 40) for the 164,297 K and M stars within 100 pc described in \S\ref{sub:train_model}. Top left panel: the precision of the predicted color. Top right panel: the precision of the predicted absolute magnitude. With our model trained only on the single stars a clear group of systems lie above the main distribution. The red dashed line lies at half of the magnitude offset expected from an equal mass system ($-2.5~log_{10}(2)/2 \approx 0.376$) and is our adopted cutoff to label stars as ``overluminous". The sources below this limit have predicted magnitudes similar to that which is measured, which we dub ``normoluminous". Center left panel: the \textit{Gaia} CMD of the normoluminous systems. Center right panel: the \textit{Gaia} CMD of the overluminous systems. Bottom left panel: the combined \textit{Gaia} CMD of the normoluminous and overluminous systems are plotted together. Bottom right panel: the \textit{Gaia} CMD of the sample colored by the fraction of stars that are overluminous. There is significant overlap in the main sequence at $M_G - A_G \sim 10$. Note that the Fraction Overluminous colormap is non-linear.}
    \label{fig:final_model}
\end{figure*}

\subsection{Final model assessment}\label{sub:train_model}

Using this ``final" model, we re-predict overluminosity for a larger set of stars with weaker constraints. We select stars from the \textit{Gaia} archive using a new query which now adds a $BP-RP$ color cut in addition to the $G-RP$ cut used earlier, but removes the RUWE constraint. The $BP-RP$ cut is added to reduce contamination near the boundary at the red end of the K dwarf sequence and removes just 384 stars. We relax the RUWE constraint because we would like to inspect whether stars with inflated astrometric errors are more likely to also be overluminous, as one would expect if astrometric errors are due to extra orbital motion. According to the simulation in \S \ref{sec:data}, we expect $\sim30$\% of the equal-mass binaries to have inflated RUWE (see Figure \ref{fig:RUWE}). Relaxing this constraint will reintroduce the unequal mass systems into the sample, but they will not be used for training our model. We inspect these suspected astrometric binaries in \S\ref{sub:Astr}.
Our query, which can be found in Appendix \ref{appendix:full_sample}, yields a sample of 183,884 stars and after performing the same BP/RP photometric excess ($C^*$) analysis as in \S\ref{sec:data} to remove stars with unreliable G-RP colors, we are left with 164,297 stars which we call the Full Sample.

We use the final (40th iteration) model from \S\ref{sub:train_model} to predict the overluminosity of this Full Sample and provide the results in Table \ref{tab:model_pred}. The predictions from this model are shown in Figure \ref{fig:final_model}. Our ``overluminous" threshold is set at half of the equal-mass binary offset ($\sim0.376$). The stars which do not exceed our threshold represent the ostensibly single systems, which we call ``normoluminous". In general, we recover binaries across the CMD and the binaries look like they are from a ``main sequence'' that has been elevated by $0.75$ mag. However, we seem to underperform in identifying binaries in two key regions: the subdwarfs ($[(G-RP)_{corrected}
,M_G - A_G]\sim[1.0, 11]$), and the very low-mass stars  ($[(G-RP)_{corrected}
,M_G - A_G]\sim[1.5, 15]$). This makes sense because our method requires a large number of stars within a location on the CMD to differentiate binary stars as outliers in spectral space; subdwarfs and very low-mass stars are inherently rare and, as a result, are sparsely populated on the main sequence. For our method to work, the overluminous binaries need to appear as outliers in a sea of single stars. 
If we inspect the locations where we see the most overluminous targets, above the main sequence most stars are labeled as binaries. As we descend into the overlapping region of the binary and single-star main sequences around $(G-RP)_{corrected} \approx 1.2$, there are regions which are 10-25\% binaries; these are the regions in the CMD where more metal-rich single stars are expected to overlap with more metal-poor binaries. Eventually we reach the lower edge of the main sequence, where there are very few overluminous binaries at all. Of the 164,297 stars in our Full Sample, we label 21,864 stars as ``overluminous" ($\sim 13\%$).

\begin{table*}[]
    \centering
    \begin{tabular}{c|c|c|c|c|c|c}
         \hline
         \textsc{source\_id} & $M_G - A_G$ & $(G-RP)_{corrected}$ & $M_G - A_G$ & $(G-RP)_{corrected}$ & Overluminous & In Banyan Sigma \\
          & (Meas.) & (Meas.) & (Pred.) & (Pred.)  &  &  \\
         \hline
         \hline
         5937146156231203072 &	9.099 & 1.049 & 9.089 & 1.044 & False & False\\
         5937154643067504512 &	10.502 & 	1.194 & 10.438 & 1.193 & False & False \\
         ... &	... & ... & ... & ... & ... & ... \\
         5961453128484376704 &	11.125 & 1.217 & 11.136 & 1.217 & False & Falses \\
         \hline
    \end{tabular}
    \caption{Catalog of overluminous stars identified with our method. Each row contains a unique \textit{Gaia} \textsc{source\_id}. The reddening for each star was calculated using the \textsc{dustmaps} \citep{dustmaps} code and the 3D dust map from \cite{edenhofer_map}. The predicted values for the absolute magnitude and color are included as well as a boolean flag as to whether the source would be considered overluminous given our requirements in \S\ref{sub:train_model}. We include a match to the BANYAN $\Sigma$ young associations catalog to flag possible overluminosity due to youth rather than binarity \citealp{banyan_XI, gagne_nearby_associations}. The catalog is 164,297 rows long so only three rows are included for the sake of brevity. The entire catalog of overluminous classifications may be downloaded at \dataset[10.5281/zenodo.18500082]{https://doi.org/10.5281/zenodo.18500082}.}
    \label{tab:model_pred}
\end{table*}

\section{Validation} \label{sec:validation}

\begin{figure*}
    \centering
    \includegraphics[width=\textwidth]{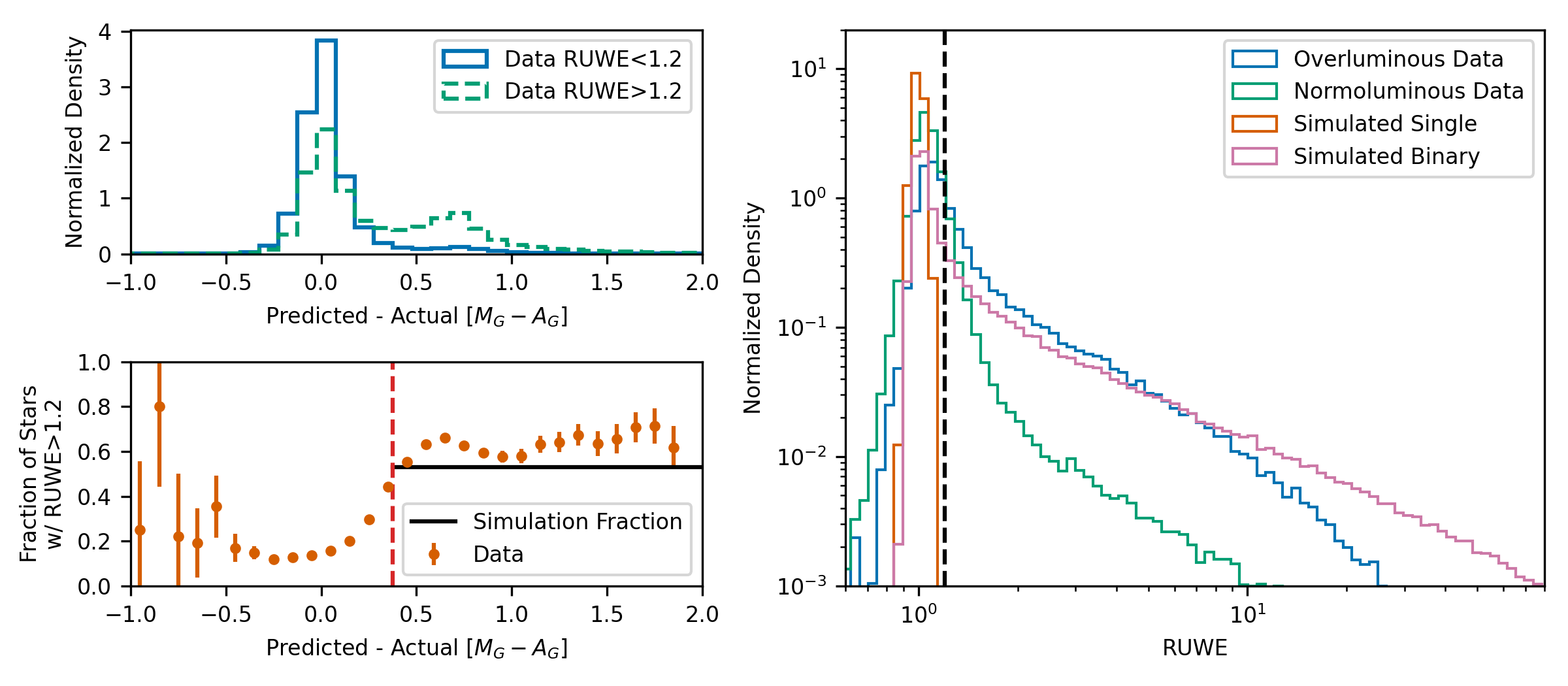}
    \caption{Correlation between RUWE and overluminosity. The top left panel shows the difference distribution of the predicted and actual \textit{Gaia} absolute G-band magnitudes. The high- and low-RUWE stars within the Full Sample defined in \S\ref{sec:method} are shown, with the high-RUWE sample showing a clear peak at the equal-mass offset ($\sim 0.75$). The bottom left panel shows the fraction of the sample which has an inflated RUWE in the data (orange). We also plot the predicted fraction of binaries (black) that lie above our magnitude cutoff and have inflated RUWE from our simulation in \S\ref{sec:data}. The right panel shows the RUWE distribution for the overluminous and normoluminous stars in the Full Sample. We also include the distribution for the single and binary systems from our simulation. }
    \label{fig:ruwe_100pc}
\end{figure*}

Earlier we discussed that there is no complete catalog of multiplicity for stars further than 20 pc. This is not to say there are not catalogs of binaries (for example the Washington Double Star Catalog or ASAS-SN Variable stars), but rather that these catalogs only contain true binary sources or ``true-positives". To use traditional machine learning metrics (e.g. True Skill Score or Heidke Skill Score) we would need a catalog of ``true-negative", bona fide, single sources. Despite our lack of labels, we can test the validity of our model by examining how many of the true-positive binaries are recovered as overluminous. In this section we compare astrometric, eclipsing, and spectroscopic methods of identifying binaries to our overluminosity method. 

\subsection{Astrometric (High RUWE) Binaries} \label{sub:Astr}

RUWE is not always inflated for equal-mass binaries, but a large RUWE is typically a signal of binarity or an atypical astrometric solution \citep{lindegren_binaries}.  We would thus expect that a significant fraction of high-RUWE stars to be flagged as overluminous with our regressor, but not all overluminous systems to have high-RUWE.

To examine how RUWE and overluminosity correlate, we separate the Full Sample into a high-RUWE subset (RUWE$>$1.2) and a low-RUWE subset (RUWE$<$1.2). We compare these two subsets in Figure \ref{fig:ruwe_100pc}.
The high-RUWE sample clearly has two distinct peaks, one centered at the zero point and the other one centered at the equal-mass binary ($M_G-A_G\simeq0.75$) offset. The low-RUWE set also has peaks at 0 and $\sim$0.75 but the latter is less pronounced. There is scatter in both distributions leading to long tails on the overluminous end. These objects are likely a mix of sources that have mispredicted magnitudes due to parallax errors (indeed the parallax error is underreported for stars with inflated RUWE \citep{elbadry_ruwe_error}) or are truly very luminous compared to similar stars, such as a young stellar object (YSO).
The underluminous side of the distribution (($M_G-A_G<0.$) is dominated by M-dwarf/white dwarf binaries and the ultra-subdwarfs stars, where our method is less effective at flagging over-luminous systems (see Fig. \ref{fig:final_model}). 
The lower left panel shows the fraction of stars with RUWE greater than 1.2 for each bin of Predicted-Actual absolute magnitude, where the errorbars are calculated assuming binomial errors of a proportion. In the region where our predicted absolute magnitude is similar to the measured value (the ``normoluminous" stars), there are few sources with high RUWE. At a larger difference between the prediction and measurement, we see that the overluminous stars have an increasing number stars with high RUWE. We also calculate from the simulation in \S\ref{sec:data} the fraction of stars which have inflated RUWE that are above the overluminosity cutoff. This fraction ($\sim0.53$) is shown as a horizontal black line in the plot and is in decent agreement with our high RUWE fraction, but our measured fraction is slightly higher. This may be due to the fact that, unlike the Full Sample, the simulation contains no single stars.

The right panel of Figure \ref{fig:ruwe_100pc} shows the RUWE distribution for four sets of stars. The overluminous, normoluminous, candidate binaries, the simulated binaries, and a set of single sources generated with \textsc{GaiaUnlimited} using the code provided by \cite{subsample_GU}. The simulated singles lie entirely below the 1.2 threshold while the simulated binaries have an extended tail towards high values. Both the normoluminous and overluminous stars have this extended feature but it is nearly an order of magnitude less pronounced in the normoluminous set. This is strong evidence that our overluminous method is recovering the astrometric binaries.

\subsection{Eclipsing Binaries} \label{sub:EB}

\begin{figure*}
    \centering
    \includegraphics[width=\textwidth]{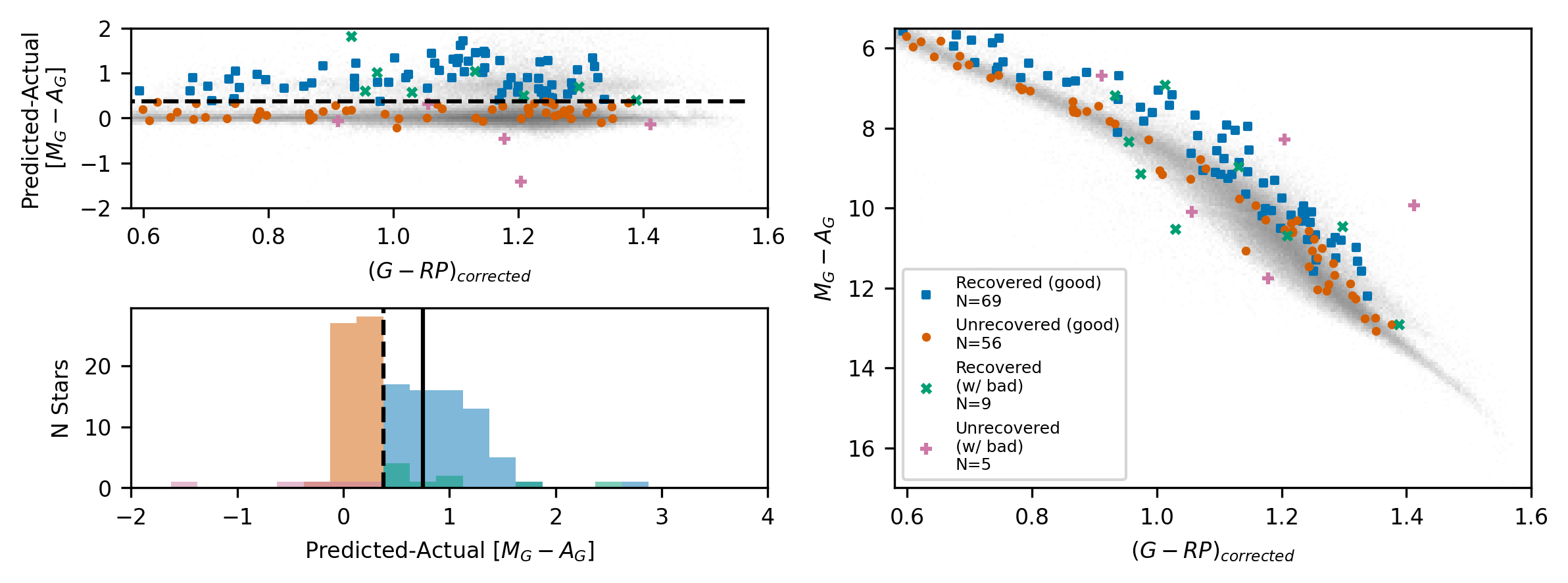}
    \caption{The AAVSO VSX eclipsing binary catalog is shown in four sets based on whether 1) they were recovered as overluminous via our method or 2) they have good photometric, spectroscopic, and astrometric data as defined in \S\ref{sec:data}. Data which pass our quality test are split into the recovered set (blue) and the unrecovered set (orange). The stars which do not pass are also split into recovered (green) and unrecovered (pink) sets. The top left panel shows the overluminosity as a function of color, with the dashed line displaying our overluminous limit. The bottom left panel displays the different samples' overluminosity as histograms with our overluminosity limit shown as a dashed line. The solid line represents the magnitude offset for an exactly equal-mass binary. The right panel shows the dereddened \textit{Gaia} CMD of the systems plotted over with the full 100 pc sample in gray.}
    \label{fig:EB_validation}
\end{figure*}

\begin{figure}
    \centering
    \includegraphics[width=0.47\textwidth]{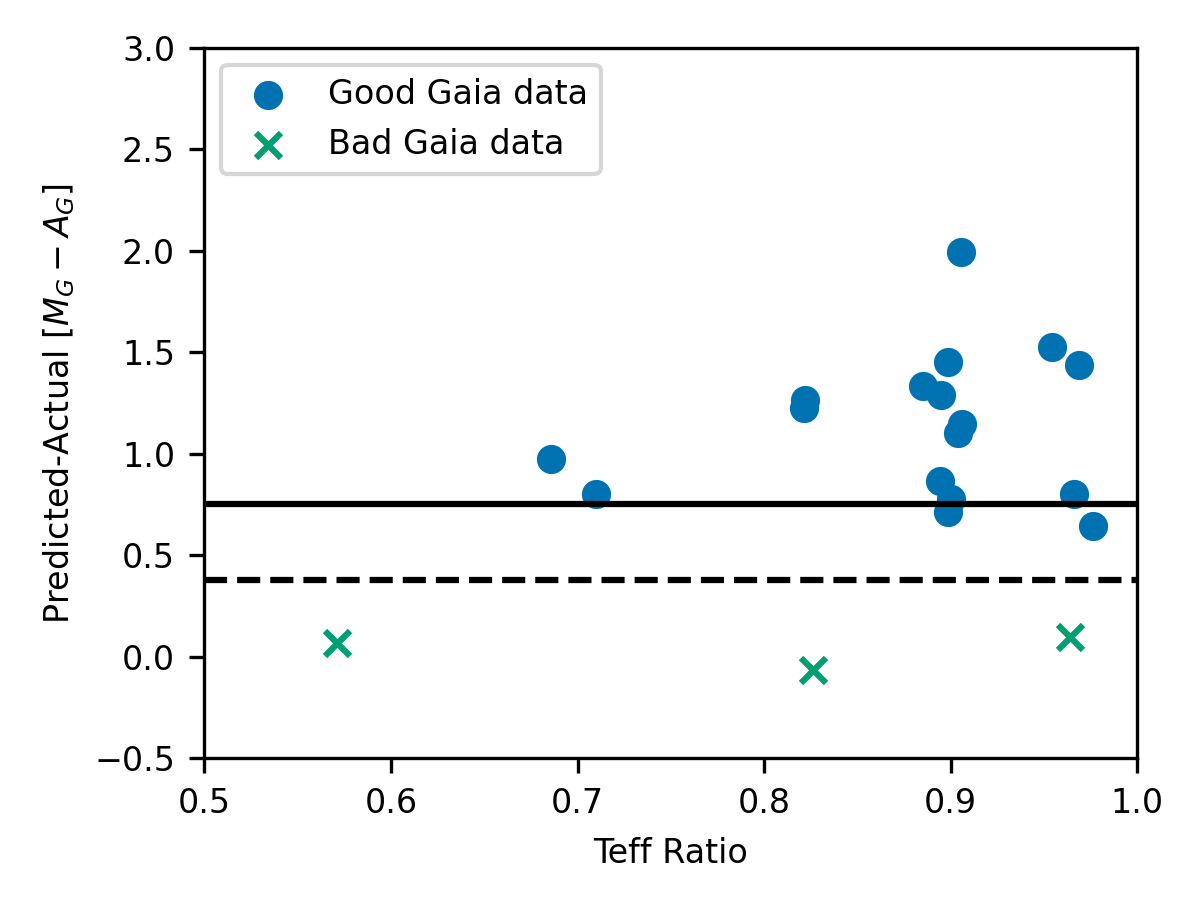}
    \caption{The relationship between the overluminosity and effective temperature ratio of the ASAS-SN Binary Stars Database. Stars are separated between those that have good (circles) and bad (crosses) \textit{Gaia} photometric and astrometric parameters. Our overluminosity cut is shown as a dashed line and the equal-mass magnitude offset is a solid line. We recover all of the binaries with good data as overluminous. }
    \label{fig:ASASSN_overluminosity}
\end{figure}

Eclipsing binaries are typically close systems and are not resolved, except in extreme circumstances. Equal-mass, eclipsing systems should all be deemed overluminous with our method unless their photometry was systematically taken during transits, which is unlikely as the \textit{Gaia} photometry typically has more than 15 observations \citep{XP_de_angeli}. To create a sample of known binary systems, we select all stars from The International Variable Star Index \citep[VSX]{VSX} from the American Association of Variable Star Observers (AAVSO) which are within 100 pc and that contain the following variable classifications (including uncertain ``$:$" and joint ``$|$",``$+$",``$/$" classifications): $E$, $EA$, $EB$, $EC$, $ED$, $ESD$, $EW$. We find 139 sources with those classifiers, 125 of which are in our Full Sample from \S\ref{sec:method} with good photometric and astrometric parameters. The 14 stars not found in
the Full Sample could have, for example, large parallax errors or blended RP transits and would not fall into our ADQL query. We perform our overluminosity analysis on the full set of eclipsing stars and show the results in Figure \ref{fig:EB_validation}. The same overluminosity cut is shown for the good and bad samples along with their \textit{Gaia} CMD. Of the 125 eclipsing binaries that have good \textit{Gaia} solutions, we recover 69 as overluminous. The 56 systems which we do not recover primarily lie within the denser part of the main sequence in the CMD and may not be equal-mass binaries. Rather, their companions may be faint and not significantly contribute to the total flux enough to make the system overluminous.

The VSX catalog does not provide a constraint on the temperature ratio, and thus we cannot examine how our recovery rate varies with mass ratio. However, a subset of 20 stars in the VSX catalog are also in the ASAS-SN Binary Stars Database \citep{asassn, rowen_asassn_binaries} which provides the best-fit effective temperature ratio to the light curve. Of these binaries, 17 systems have good \textit{Gaia} data and we recover them all as overluminous. We show the relationship between their overluminosity and the effective temperature ratio in Figure \ref{fig:ASASSN_overluminosity}. These sources span temperature ratios from $\sim 0.7$ to $1.0$, with most having ratios above $0.9$. It would be surprising if we were actually seeing overluminosity at temperature ratios below $0.9$, so we suspect that these systems may have contaminated light curves or poor fits.

It would be possible to perform a further assessment of the eclipsing binaries in the \textit{Catalog of Systems} using the lightcurves from the TESS mission \citep{TESS}. However, studies that search for variability in the TESS data have trouble attributing that variability to a specific star because the TESS pixels are $\sim 21$ arcsec wide. Analyzing variability usually has to be done on a case-by-case basis (e.g., \citet{kar_variability}) and we consider such an effort beyond the scope of this work.

\subsection{Spectroscopic Binaries} \label{sub:SB}
\begin{figure*}
    \centering
    \includegraphics[width=\textwidth]{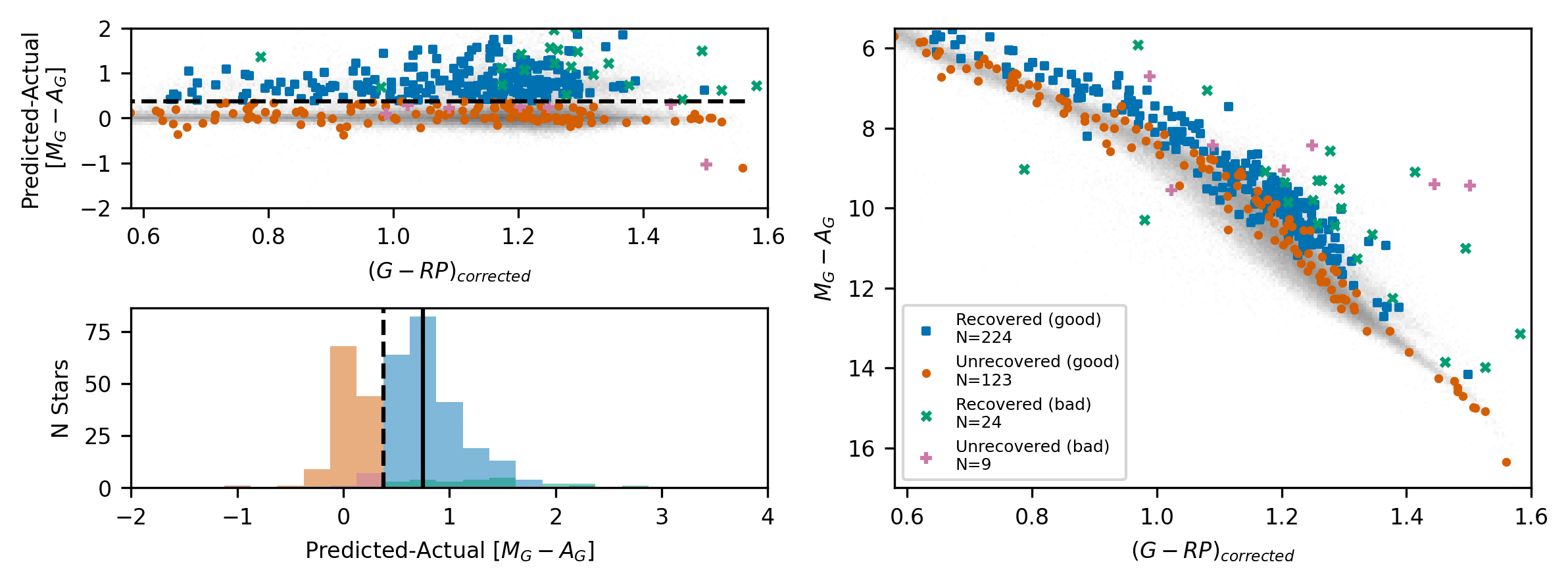}
    \caption{The binary sample identified using the $\Delta RV_{max}$ method. The plots are displayed using the same convention as in Fig. \ref{fig:EB_validation}.}
    \label{fig:deltaRVmax}
\end{figure*}

\begin{figure*}
    \centering
    \includegraphics[width=\textwidth]{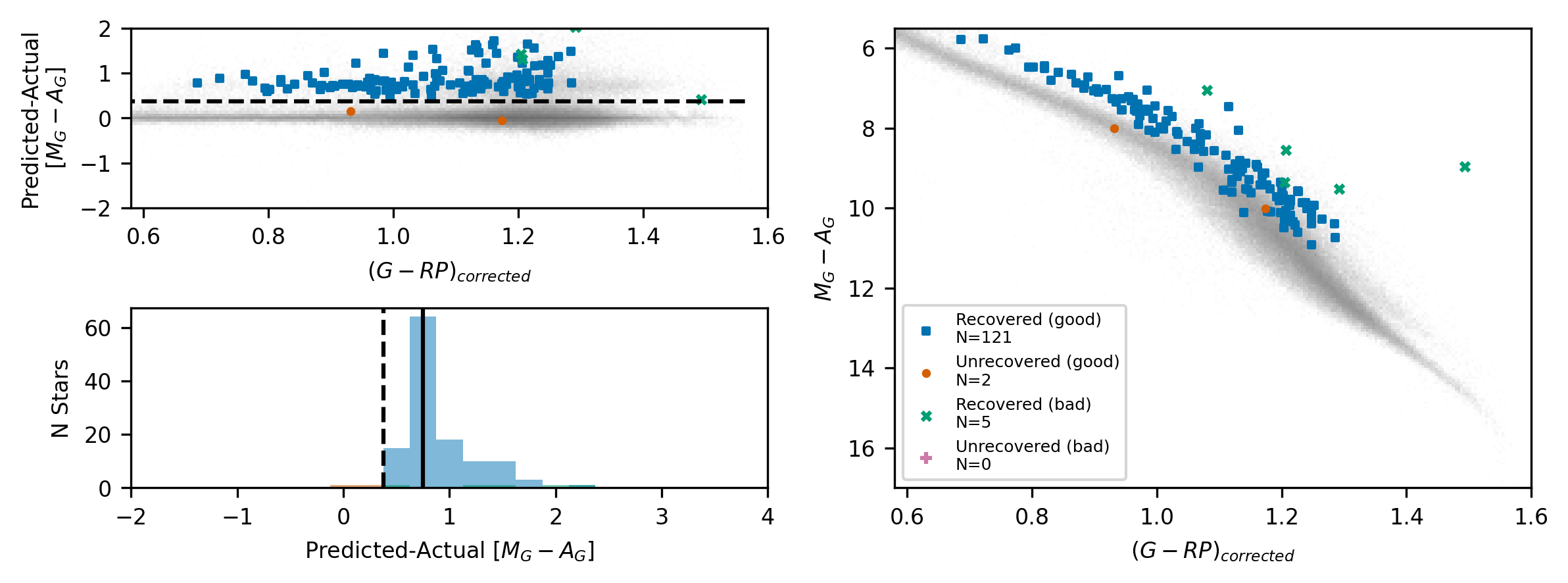}
    \caption{The SB2 binary catalog from \citep{kounkel_SB2}. The plots are displayed using the same convention as in Fig. \ref{fig:EB_validation}.}
    \label{fig:kounkelSB2}
\end{figure*}

Spectroscopic binaries are also typically close systems and will be unresolved in the XP spectra. We utilize two methods to identify spectroscopic binaries from the APOGEE DR17 catalog \citep{apogee_DR17}: RV variability and double-lined detection.

For RV variability, we follow the method of \cite{delta_rv_max}. First, we find the individual RV measurements from the DR17 \textsc{allVisit} file. Then, for each source we calculate the maximum difference in RV ($\Delta RV_{max}$). \cite{delta_rv_max} found that populations with $\Delta RV_{max}>3~km/s$ were dominated by single- and double-lined spectroscopic binaries. We adopt this cut to select a sample of ``true" binaries and perform our overluminosity method. The results are shown in Figure \ref{fig:deltaRVmax} with the same split between ``good" and ``bad" \textit{Gaia} parameters as in \S\ref{sub:EB}. Of the 347 binaries identified by RV variation, we recover 224 ($~65\%$) as binaries. As with the eclipsing binaries, we see that we are best at recovering sources above the main sequence and there is a mix of recovered and unrecovered systems in the denser part of the main sequence. These systems may be SB1 binaries with an undetectable faint companion, and thus not recoverable by our overluminosity method. The recovered systems in the dense are are likely metal-poor, near equal-mass systems that are raised to be more luminous. Without fitting the time series RVs, it is not possible to constrain the mass ratio.

We utilize the double-lined spectroscopic binary catalog from \cite{kounkel_SB2} to create a well-defined, equal-mass sample of SB2 systems. This study examined the cross-correlation function (CCF) of each of the APOGEE DR16 and DR17 stars with the best fit template. In this case we can assemble a set of true, near equal-mass systems, by selecting K and M dwarfs sources which have only 2 components and well defined peaks ($N=2$ with $Flag~4=True$) within 100 pc and with XP spectra. We perform the overluminosity analysis and show the results in Figure \ref{fig:kounkelSB2}. We recover 121 of the 123 ($\sim98\%$) spectroscopic binaries as overluminous, which suggests that our method is most efficient in identifying equal-mass systems, but compared to the $\Delta RV_{max}$ method we are not robust in identifying every binary. 

\begin{figure}
    \centering
    \includegraphics[width=0.47\textwidth]{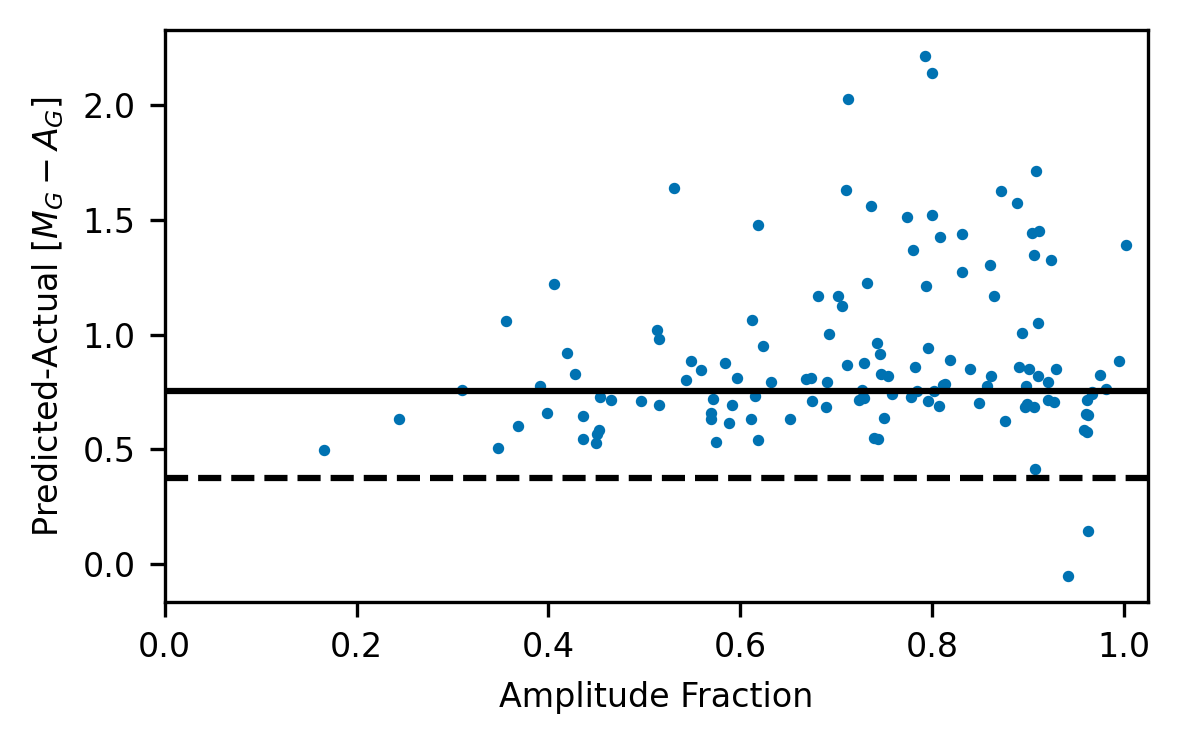}
    \caption{Overluminosity of the SB2 binaries from \cite{kounkel_SB2} are shown against the amplitude fraction of the cross-correlation peaks. Our overluminosity cut is shown as a dashed line and the equal-mass magnitude offset is a solid line. The fraction will be close to 1.0 when the best-fit template matches both stars in the measured spectrum, making it a likely equal-mass system. It can be seen that our overluminosity method performs well across the range of amplitude fractions.}
    \label{fig:SB2_amp_frac}
\end{figure}

We also examine how near these systems are to being equal-mass by taking the ratio of the smaller amplitude signal in a star's CCF to the larger signal. We show this amplitude fraction in Figure \ref{fig:SB2_amp_frac}. Stars with amplitude fractions closer to 1 will have an SED closer to two duplicated stars with the best-fit, cross-correlated model describing both stars equally well. Fractions closer to 0 will have a small second peak, signaling that the secondary star is both dimmer and has a significantly different SED. It can be seen that we perform equally well across amplitude fractions, but there is a slight downward trend below the overluminosity cutoff for lower amplitude fractions.

\section{Catalog}\label{sec:catalog}

Our method of identifying unresolved binaries is most sensitive to close ($\lesssim 20$ AU), equal-mass systems because of the \textit{Gaia} resolution limit of $\sim0.2$ arcsec for visual binary stars\citep{res_limit}. However, in this section we aim to aggregate a more complete census of binaries in \textit{Gaia} in the Solar Neighborhood ($d<100~pc$). We supplement our identification of probable unresolved binaries for the low-mass stars with other recent works which identify wide binaries in co-moving pairs and visual binaries at short separations. By combining these catalogs, we assemble a more comprehensive census of multiple stellar systems across a wide range of separations.

\subsection{Multiplicity Catalogs}\label{subsec:mult_cats}

We collate a list of all multiple sources within 100 pc from five sources. The first two are (1) the wide binary catalog from \cite{elbadry_WB} and (2) the resolved triple systems from \cite{shariat_triples}. Both of these methods search for gravitationally bound, common proper motions stars in the \textit{Gaia} DR3 catalog. These catalogs utilize simple cuts on projected separation, parallax difference, and proper motion difference constrained by Keplerian motion from the binary orbit to create their catalog of widely separated systems. We limit the \cite{elbadry_WB} catalog by restricting pairs to those with chance alignment probability $R<0.1$. 

We complement these two catalogs with (3) the SUPERWIDE wide binary catalog \citep{superwide}, which identifies common proper motion binaries using a Bayesian approach for stars with proper motions $>40~mas$ identified in the SUPERBLINK survey. This search is optimized to find nearby wide systems, and includes 3,116 \textsc{source\_id} pairs not listed in \cite{elbadry_WB} or \cite{shariat_triples}.

An additional set of binaries we include are (4) the visual binaries identified by \cite{medan_visual_binaries}. These are systems which have very low projected separations ($<2.5$ arcsec). Stars below this separation are resolved in \textit{Gaia} DR3, but one of the components often lacks an astrometric solution (no parallax or proper motion); the components are also usually unresolved in the 2MASS catalog. \citet{medan_visual_binaries} developed an empirical model to identify systems that are likely to be gravitationally bound and have projected separations $< 30~$ AU. We keep only the most likely visual binaries from \citet{medan_visual_binaries}  by selecting those with a contamination factor $L<0.00193$ and contamination rate $C_{ideal}<0.1$. This catalog includes 2,662 \textsc{source\_id} pairs not listed in \cite{elbadry_WB} or \cite{shariat_triples}.

The final source we use is (5) the \textit{Gaia} non-single star (NSS) catalogs \citep{gaia_NSS}. These catalogs summarize the methods which the \textit{Gaia} Collaboration developed to search for signs of binarity in the astrometry, photometry, and spectroscopy in DR3. These include searching for two-body orbits, astrometric or spectroscopic accelerations, and variable binaries.

\subsection{Creating a \textit{Catalog of Systems}}\label{subsec:syst_cat}

Our goal here is to assemble a \textit{Catalog of Systems} within 100 parsecs based on the \textit{Gaia} DR3 catalog. Each \textsc{source\_id} will be treated as a unique object which can be associated with other \textsc{source\_id}s in the source catalogs above. At the end, we will have a table where each row is a unique stellar system with up to 5 \textsc{source\_id}s associated within the system.

We begin with the \cite{shariat_triples} triples. 
Then we loop through the resolved binary catalogs \citep{elbadry_WB, superwide, medan_visual_binaries} and find any possible association between one \textsc{source\_id} and another. There are three possibilities at this stage:
\begin{enumerate}
    \item The binary association already exists in the catalog
    \item One \textsc{source\_id} is in the catalog, but is associated with different \textsc{source\_id}
    \item Neither \textsc{source\_id} exists in the catalog
\end{enumerate}
For case (1) there is nothing new to be added to the catalog. In case (2), we append the new \textsc{source\_id} to the existing system.\footnote{For example, one \textsc{source\_id} is associated with different \textsc{source\_id}s in \cite{medan_visual_binaries} and \cite{elbadry_WB}; at this stage there will be three separate \textsc{source\_id}s from the two different associations.}. Finally, in case (3) we add the two \textsc{source\_id}s as a new row. After looping through the binary catalogs, we add every source from \textit{Gaia} DR3 which has $\varpi>10$ to include all stars within 100 pc, including single-star systems. The resulting table then has one row per unique stellar system with columns for each unique \textsc{source\_id} in that system. Many systems only contain a single \textit{Gaia} \textsc{source\_id} while others may contain up to 5 \textsc{source\_id}s. We sort the \textsc{source\_id}s by their \textit{Gaia} G-band magnitude with the brightest value first. We list each \textsc{source\_id} in Table \ref{tab:catalog} with a ``\_n'' added to the end. Here, ``n'' refers to the numbered place that each \textsc{source\_id} is placed in each row (or system) and can be 1, 2, 3, 4, or 5. Each column in the table that ends in ``\_n'' is repeated five times for each possible \textsc{source\_id} in the system. To trace each \textsc{source\_id}'s relationship to the catalogs, we provide the column \textsc{catalogs\_n} which is an 8-bit integer. The first five bits signal membership in the supporting catalogs. The bit order is (1) \cite{shariat_triples}, (2) \cite{elbadry_WB}, (3) \cite{medan_visual_binaries}, (4) \cite{superwide}, and finally (5) the overluminosity catalog defined in \S\ref{sub:train_model} (see Table \ref{tab:model_pred}).

\subsection{Unresolved Multiplicity}\label{subsec:unre_mult}

Some of the \textsc{source\_id}s in each system still show signs of unresolved multiplicity. This may be from our analysis of overluminosity or from one of the indicators in \textit{Gaia} DR3.

We first label each \textsc{source\_id} that is found to be overluminous from our analysis in \ref{sec:method} as \textsc{True} in  the ``ol\_n" column in Table \ref{tab:catalog}.
However, since higher-order systems may be in moving groups and therefore inherently young, these objects might be overluminous YSOs rather than unresolved binaries. Furthermore, these moving group members may be co-moving because they originate from the same star-forming cloud, rather than being gravitationally bound. To account for the young objects, we match the \textit{Catalog of Systems} to the young associations identified with BANYAN $\Sigma$ and listed in the MOCAdb database\footnote{https://mocadb.ca} (\citealp{banyan_XI, gagne_nearby_associations}; Gagn\'e et al., ApJS, under review). Stars with $>50\%$ probability of being in an association are labeled with the ``in\_banyan\_sigma\_n'' column. Our identification of the stars being ``overluminous" in those cases should not be treated as a sign of multiplicity when the source is in a young association, but likely as a sign of a larger stellar radius. MOCAdb also provides matches to other young associations from the literature. We provide the column ``lit\_membership\_references\_n'' which lists the references that claim that the \textsc{source\_id} may be a member of a young association. However, some of these sources are not flagged by BANYAN $\Sigma$. These catalogs are heterogeneous and can vary in precision so, while we provide the list of references, we do not include them in any further analysis and use BANYAN $\Sigma$ as our indicator of youth.

We also check for signs of binarity from the \textit{Gaia} catalog using the criteria defined in Table 3 of \cite{carmenes_multiplicity}. These criteria provide limits on the astrometry and radial velocities using \textit{Gaia} parameters which signal multiplicity. If a source is triggered with any of these criteria, the ``carmenes\_flag\_n'' column is marked as \textsc{True}. We include in the catalog each pertinent column from \textit{Gaia} DR3 for each of these criteria. We also cross-match each \textsc{source\_id} to the \textit{Gaia} NSS catalogs to recover which type of binary a source might be. We include their designation in columns beginning with ``nss\_'' in Table \ref{tab:catalog}. We also provide the values for the 2MASS \citep{2mass} and AllWISE \citep{allwise} photometry using the cross-match provided in the \textit{Gaia} DR3 archive.

The summary of each column in our catalog is shown in Table \ref{tab:catalog}. For convenience, we include \textit{Gaia} astrometry, photometry, radial velocities, and other relevant parameters. Finally, we count the total number of members in each system as the number of \textsc{source\_id}s added with the possible unresolved sources. That is, for each overluminous component which is not in a young association (\textsc{ol\_n} == \textsc{True} and \textsc{in\_banyan\_sigma\_n} == \textsc{False}), we add a member to the total count for the system. We also include those sources that signal an unresolved companion in \cite{carmenes_multiplicity} as long as they are not overluminous, so as to not double count an unresolved companion (\textsc{carmenes\_flag\_n}==\textsc{True} and \textsc{ol\_n}==\textsc{False}). The sum of the possible unresolved components is included in the ``\textsc{n\_components}'' column. 

\begin{figure}
    \centering
    \includegraphics[width=0.47\textwidth]{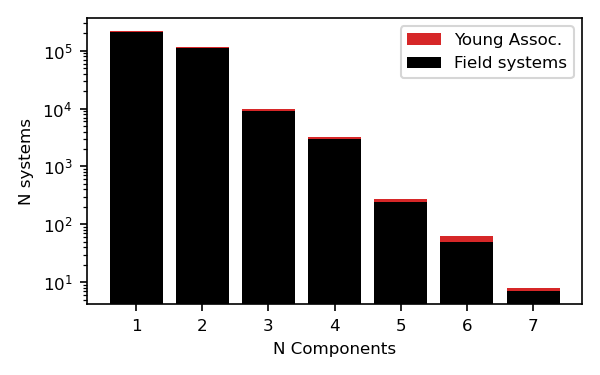}
    \caption{Total number of systems identified as a function of the number of components N in the system. Systems which are suspected members of young associations are added in red.}
    \label{fig:catalog_hist}
\end{figure}

\begin{figure*}
    \centering
    \includegraphics[width=\textwidth]{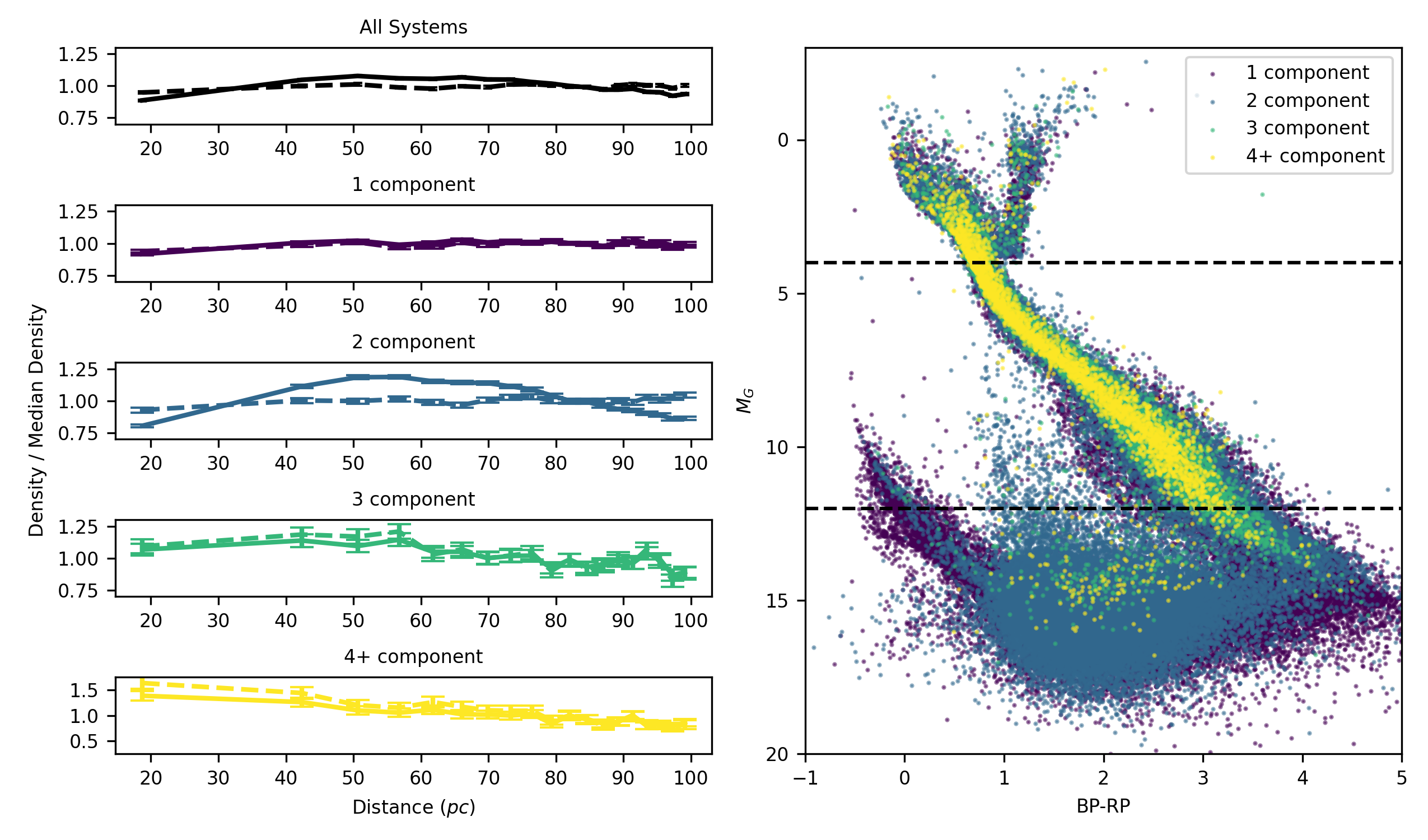}
    \caption{The density of the \textit{Catalog of Systems} and its \textit{Gaia} CMD. Left panel: the density normalized by its median for bins of equal-volume. The density for all systems is shown in the top panel and then split into the number of components in the lower panels. The solid lines mark the normalized density for all systems which have each number of components while the dashed lines show the normalized density only for sources with absolute magnitude between 4 and 12 (shown as black dashed lines in the right panel). Right panel: \textit{Gaia} CMD of the primary component of each of these systems. Points are colored by the number of components in each system. There are many systems which lie between the main sequence and white dwarf sequence, which are a set of known spurious sources. We elect not to remove these sources from the catalog so that a user can produce their own cut.}
    \label{fig:catalog_hr}
\end{figure*}

In total, we find 347,440 systems of up to 5 \textit{Gaia} \textsc{source\_id}s and with up to 7 components. We show the distribution of total systems by number of component in Figure \ref{fig:catalog_hist}, with the suspected members of young associations highlighted. To investigate possible systematic issues with the catalog we find the density of systems across equal-volume shells in Figure \ref{fig:catalog_hr}. The densities for all systems and for each number of components, one through four or greater, are normalized to their median value. There is a dearth of systems for all numbers of components in the most nearby bin due to \textit{Gaia}'s brightness limit. Stars bright than $G\sim3$ saturate \textit{Gaia}'s CCD and a parallax cannot be found \citep{Gaia_DR3}. For the systems with two components, there is an abundance of systems around 50-60 pc. This is not a real effect, but is instead a population of known spurious sources. We can see these systems in the the \textit{Gaia} CMD in  Figure \ref{fig:catalog_hr}. There are many systems that lie between the main sequence and the white dwarf sequence, which are known to have inflated parallax due to issues resolving close pairs \citep{gcns, dr2_summary}. We choose not to remove these systems so as to provide a more holistic set of \textsc{source\_id}s which can be pruned depending on the desires of the user. However, to continue investigating the systematic issues of our catalog we repeat the exercise of finding the normalized density for sources with absolute magnitude ($M_G$) between 4 and 12. We now largely avoid the spurious sources and show that the crest of systems with 2 components is removed. There is still a slight upward trend, signaling that we may be missing binaries at shorter distances. This is probable since there is a known population of missing binaries that have short separations and near-equal magnitudes \citep{gaia_hole_tokovinin}.

\begin{table*}[]
    \centering
    \begin{tabular}{l|c|c|r}
        \hline
         Column Name & Data Type & Lite & Description \\
         \hline \hline
         source\_id\_n & Integer & $\checkmark$ & The \textit{Gaia} DR3 source identifier \\
         catalogs\_n & Integer & $\checkmark$ & An 8-bit integer which identifies catalog membership (see \S\ref{subsec:syst_cat})  \\
         ol\_n    & Bool & $\checkmark$ & Whether the source is considered overluminous \\
         in\_banyan\_sigma\_n     & Bool & $\checkmark$ &  Whether this source is in BANYAN $\Sigma$\\
         lit\_membership\_references\_n     & String & $\checkmark$ & Reference that this \textsc{source\_id} belongs to according to MOCAdb\\
         ra\_n    & Float & & The \textit{Gaia} DR3 Right Ascension (J2016)\\
         dec\_n  & Float & &  The \textit{Gaia} DR3 Declination (J2016)\\
         pmra\_n & Float & &  The \textit{Gaia} DR3 Right Ascension proper motion\\
         pmdec\_n    & Float & &  The \textit{Gaia} DR3 Declination proper motion\\
         parallax\_n & Float & &  the \textit{Gaia} DR3 parallax\\
         radial\_velocity\_n  & Float & &  The \textit{Gaia} DR3 radial velocity \\
         radial\_velocity\_error\_n    & Float & &  The \textit{Gaia} DR3 radial velocity error\\
         phot\_g\_mean\_mag\_n  & Float & &  The \textit{Gaia} DR3 G-band magnitude\\
         phot\_rp\_mean\_mag\_n & Float & &  The \textit{Gaia} DR3 RP-band magnitude \\
         phot\_bp\_mean\_mag\_n & Float & &  The \textit{Gaia} DR3 BP-band magnitude\\
         phot\_bp\_rp\_excess\_factor\_n & Float & &  The \textit{Gaia} DR3 BP/RP excess factor\\
         ruwe\_n & Float & &  The \textit{Gaia} DR3 RUWE\\
         ipd\_frac\_multi\_peak\_n  & Integer & &  The \textit{Gaia} fraction of astrometric multiple peaks\\
         ipd\_frac\_odd\_win\_n & Integer & &  The \textit{Gaia} DR3 fraction of odd astrometric windows \\
         ipd\_gof\_harmonic\_amplitude\_n   & Float & &  The \textit{Gaia} DR3 amplitude of the IPD goodness of fit\\
         duplicated\_source\_n   & Bool  & &  The \textit{Gaia} DR3 duplicated source flag\\
         rv\_chisq\_pvalue\_n  & Float & &  The \textit{Gaia} DR3 radial velocity P value\\
         rv\_renormalised\_gof\_n  & Float & &  The \textit{Gaia} DR3 radial velocity renormalized goodness of fit\\
         rv\_nb\_transits\_n  & Float  & &  The \textit{Gaia} DR3 number of radial velocity transits\\
         non\_single\_star\_n  & Integer & &  \textit{Gaia} DR3 3 bit flag for non-single star (NSS) catalog membership \\
         nss\_acceleration\_astro\_solution\_type\_n & String & & \textit{Gaia} DR3 non-single acceleration designation\\
         nss\_non\_linear\_spectro\_solution\_type\_n & String & & \textit{Gaia} DR3 non-single spectroscopic binary designation\\
         nss\_two\_body\_orbit\_solution\_type\_n & String & & \textit{Gaia} DR3 non-single star orbital two-body designation\\
         nss\_vim\_fl\_solution\_type\_n & String & &  \textit{Gaia} DR3 non-single Variability Induced Mover (VIM) designation\\
         j\_m\_n & Float & &  The 2MASS J-band photometry \\
         h\_m\_n & Float & &  The 2MASS H-band photometry \\
         ks\_m\_n & Float & &  The 2MASS Ks-band photometry \\
         w1mpro\_n & Float & &  The AllWISE W1-Band profile photometry \\
         w2mpro\_n & Float & &  The AllWISE W2-Band profile photometry \\
         w3mpro\_n & Float & &  The AllWISE W3-Band profile photometry \\
         w4mpro\_n & Float & &  The AllWISE W4-Band profile photometry \\
         ra\_j2000\_n & Float & &  The proper motion propagated \textit{Gaia} DR3 coordinates (J2000) \\
         dec\_j2000\_n    & Float & &  The proper motion propagated \textit{Gaia} DR3 coordinates (J2000) \\
         l\_j2000\_n & Float & &  The Galactic longitude calculated from the J2000 coordinates \\
         b\_j2000\_n & Float & &  The Galactic latitude calculated from the J2000 coordinates \\
         carmenes\_flag\_n    & Bool & $\checkmark$ &   \cite{carmenes_multiplicity} \textit{Gaia} DR3 multiplicity flag \\
         n\_components    & Integer & $\checkmark$ &  The total number of components in this systems\\
         \hline
    \end{tabular}
    \caption{The columns in the \textit{Catalog of Systems} are shown here. Since there may be multiple components, columns are repeated for each unique \textit{Gaia} \textsc{source\_id} in the system. We represent each unique column with an ``$\_n$'' which can be one of each each of the five components ($n\in{1,2,3,4,5}$). The catalog is 206 columns and/ 347,440 rows. For convenience, we also offer a ``lite'' version of the catalog which does not include the supplemental \textit{Gaia} columns. Both the full and lite versions of this table are available for download at \dataset[10.5281/zenodo.18500082]{https://doi.org/10.5281/zenodo.18500082}.}
    \label{tab:catalog}
\end{table*}

\section{Results}\label{sec:results}

\begin{figure*}
    \centering
    \includegraphics[width=\textwidth]{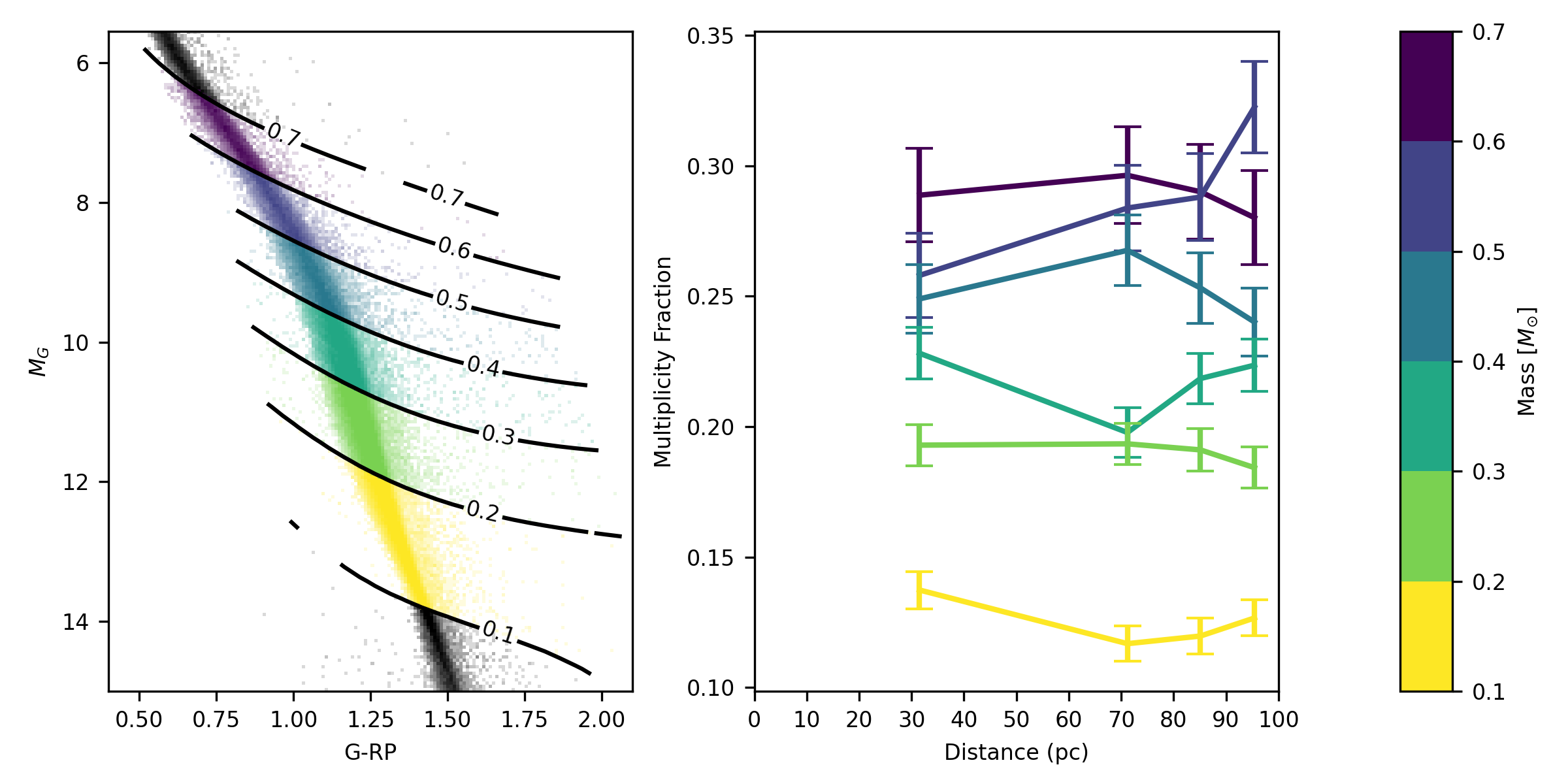}
    \caption{The multiplicity fraction of K and M dwarf stars in the \textit{Catalog of Systems}. Left panel: the \textit{Gaia} CMD of primary, low-mass star in the \textit{Catalog of Systems} (with the overluminous stars shifted by $\Delta M_G = 2.5 log_{10}|2|$). Overlaid on the CMD are the contours of equal mass derived from the relation in \S\ref{sec:results} \cite{mann_mass}. Right panel: the multiplicity fraction for each mass bin across equal-volume distances. Each fraction is colored by its mass bin.}
    \label{fig:catalog_multiplicity_frac}
\end{figure*}

One interesting constraint that this updated {\em Catalog of Systems} can provide is an approximation of the multiplicity rate as a function of mass on the lower main sequence. For this, we first need to estimate masses for our stars. \citet{mann_mass} demonstrated that for stars with metallicity similar to the Sun a tight correlation exists between a star's absolute 2MASS $K_s$ magnitude and mass. While it is possible to find the mass for every star directly from the cross-matched 2MASS photometry, there are many systems in the \textit{Catalog} which do not have a matched source or are blended. To circumvent this issue we project the values of mass onto the \textit{Gaia} color-magnitude diagram. This was done by finding the mass for every primary ($n=1$) source in the {\em Catalog of Systems} with a $K_s$ magnitude. The mass is determined using Equation 4 from \cite{mann_mass} with 5 coefficients. If the source is overluminous according to our method, then we subtract an equal-mass magnitude offset ($\Delta M_{K_s} = 2.5 log_{10}|2|$). For systems where the primary does not have a parallax, we use the parallax from the secondary star ($n=2$) - which happens to exist in all cases. We restrict our analysis to stars with masses between $0.1$ and $0.7~M_{\odot}$ where the mass-magnitude relation of \citet{mann_mass} is best constrained. We exclude the white dwarfs by requiring all stars to satisfy the inequality $M_G < 10* (G-RP) +5$. Next, we find the median mass value across the \textit{Gaia} $M_G$ versus $G-RP$ color-magnitude diagram in bins 0.25 and 0.05, respectively (also shifting the overluminous stars in $M_G$). Finally, we interpolate the median values using cubic, bivariate B-splines. This produces a functional form which takes a star's $M_G$ and $G-RP$ position and predicts a mass. By using the median value in each bin on the \textit{Gaia} CMD, we reduce issues introduced by the differences in spatial resolution. The resulting contours of mass from this method are displayed in Figure \ref{fig:catalog_multiplicity_frac}. The contours fall diagonally across the main sequence, which is expected given that we are deducing mass from the infrared photometry. The blue end is where the subdwarfs lie, which is not well constrained by the relation in \cite{mann_mass}. At the red end, we see stars with contaminated or blended RP photometry. These systems may have mispredicted masses but they are few enough to not drastically effect our measurement of multiplicity fractions. 

The \textit{Catalog of Systems} aggregates several methods of identifying multiple systems and suffers biases due to resolution and magnitude limits of the \textit{Gaia} instrument. To reveal possible biases induced by \textit{Gaia}'s resolution, we calculate the multiplicity fraction of the stars across equal-volume bins in each mass bin. The results are shown in Figure \ref{fig:catalog_multiplicity_frac}. The mass bins, distance bins, and multiplicity fractions are listed in Table \ref{tab:multiplicity}. The error for each fraction is calculated assuming the binomial error of a proportion. Most multiplicity fraction estimates are consistent across the distance bins, except for the $M\in[0.5,0.6]~M_{\odot}$ bin where we see a dramatic rise as the distance increases. This may be because the {\em Catalog of Systems} does not contain binaries in three key regimes. The first is the unequal mass systems where the secondary is not luminous enough to add to the total flux and be detected as overluminous; we already discussed these systems in depth in \S\ref{sec:validation}. The second regime is the binary sources that do not have \textit{Gaia} astrometry because they lie in the ``Gaia hole''. \cite{gaia_hole_tokovinin} shows that these systems are typically equal-mass with angular separations slightly larger than \textit{Gaia}'s resolution, causing the astrometric solution to fail. The third regime is the undetected binaries which exist in \textit{Gaia} DR3, but do not have an XP spectrum and would thus not be identified as overluminous with our method. The lower magnitude limit for sources with a reported XP spectrum is $G<17.65$. In our 100 parsec sample, this very likely causes the lowest-mass bin ($M_G>12$) to have a lower multiplicity fraction. Due to these biases, our multiplicity fraction is best thought of as a {\em lower limit} rather than a direct estimate.

\begin{figure*}
    \centering
    \includegraphics[width=\textwidth]{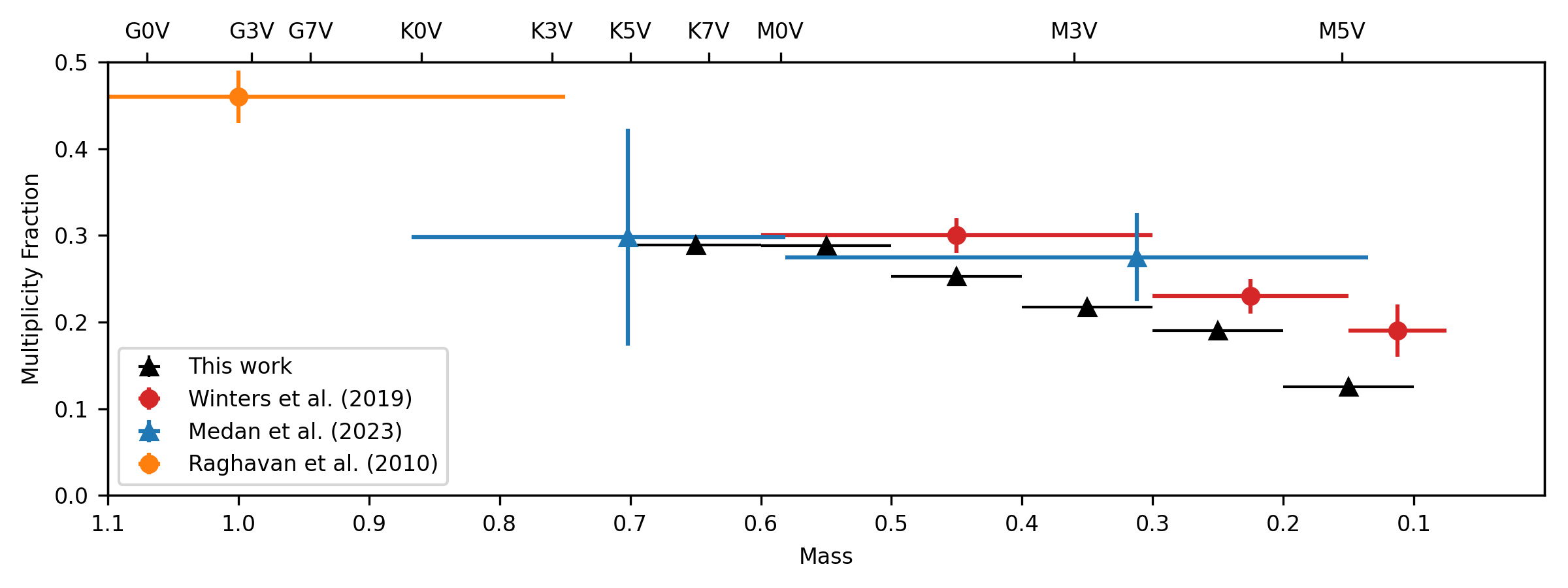}
    \caption{A comparison between our multiplicity fraction and other work. In black are the total multiplicity fraction for each mass bin. Included in the plot are multiplicity fractions from \cite{winters_m_multiplicity} (in red), \cite{stop_cop_city} (in orange), and \cite{medan_visual_binaries} (in blue). Points represented as dots are measured values whereas upward arrows represent lower limits. The horizontal bars represent the width of the bins and not an error.}
    \label{fig:catalog_multiplicity_mass}
\end{figure*}

Our multiplicity fraction for each mass bin is compared to results from previous work in Figure \ref{fig:catalog_multiplicity_mass}. Our estimates are represented as upward arrows to indicate that they are lower limits on each fraction. We display the results from \cite{stop_cop_city} (for Solar-type stars in orange),  \cite{winters_m_multiplicity} (for M-type stars in red) and \cite{medan_visual_binaries} (in blue). The first two papers infer the multiplicity from complete samples and are represented as solid points in the plot. The \citet{medan_visual_binaries} estimates are complete only for the range of physical separations $s \in [30,500]~au$ and thus also represent lower limits on the total multiplicity fraction. Our own lower bounds are consistent with the results from all these works and reproduce the same trend of multiplicity fractions decreasing with system mass. We note that N-body simulations of low-mass star formation however predict much lower multiplicity fractions than our lower limits suggest, signaling a greater need for understanding disc fragmentation at low masses \citep{starforge_binarity}.

\begin{table*}[]
    \centering
    \begin{tabular}{c|c|c|c|c|c}
         \hline
          Mass [$M_{\odot}$]& 0-63 pc & 63-79 pc & 79-91 pc & 91-100 pc & 0-100 pc\\
          \hline \hline
         $M \in [0.1, 0.2]$ & $0.137 \pm 0.007$ &
$0.117 \pm 0.007$ &
$0.120 \pm 0.007$ &
$0.127 \pm 0.007$ & $0.125 \pm 0.003$ \\
         $M \in [0.2, 0.3]$ & $0.193 \pm 0.008$ &
$0.193 \pm 0.008$ &
$0.191 \pm 0.008$ &
$0.184 \pm 0.008$ & $0.191 \pm 0.004$ \\
         $M \in [0.3, 0.4]$ & $0.228 \pm 0.010$ &
$0.198 \pm 0.009$ &
$0.218 \pm 0.010$ &
$0.224 \pm 0.010$ & $0.217 \pm 0.005$ \\
         $M \in [0.4, 0.5]$ & $0.249 \pm 0.013$ &
$0.268 \pm 0.014$ &
$0.253 \pm 0.014$ &
$0.240 \pm 0.013$ & $0.253 \pm 0.007$ \\
         $M \in [0.5, 0.6]$ & $0.258 \pm 0.016$ &
$0.284 \pm 0.016$ &
$0.288 \pm 0.017$ &
$0.322 \pm 0.018$ & $0.288 \pm 0.008$ \\
         $M \in [0.6, 0.7]$ & $0.289 \pm 0.018$ &
$0.296 \pm 0.019$ &
$0.290 \pm 0.018$ &
$0.280 \pm 0.018$ & $0.289 \pm 0.009$ \\
         \hline
    \end{tabular}
    \caption{Multiplicity fraction for different absolute magnitudes and equal-volume distance bins. The final column shows the multiplicity fraction over the full distance range. Absolute magnitudes which are significantly biased by systematic issues are marked with an asterisk and discussed in \S\ref{sec:results}. Faint companions are not detected in our search for overluminosity and may be missed in \textit{Gaia}, thus these multiplicity fractions should be considered lower limits.}
    \label{tab:multiplicity}
\end{table*}

\section{Summary and Discussion}\label{sec:summary}

In this work we identify 21,864 low-mass stars as overluminous binary candidates. We do this by training a Random Forest Regression model to predict a star's location on the color-magnitude diagram given its \textit{Gaia} XP spectrum. The binaries appear more luminous than the model's prediction. 

We begin the paper by constructing the Regression Sample. In \S\ref{sec:data} while describing our selection, we simulate a set of low-mass binaries following a physically motivated mass ratio distribution. After using the \textsc{GaiaUnlimited} software, we show that a selection based on the RUWE parameter leaves out more than half of the equal-mass binaries. Since there is an excess of ``twin'' binaries for the low-mass stars, we highlight that a selection based on \textit{Gaia}'s astrometric error will not result in a sample of single systems. 

After selecting our Regression Sample, we iteratively train regression models to remove binary stars. For each model, sources are predicted to be significantly fainter than their counterparts from their XP spectra. We converge on a model trained on the effectively single stars because the spectra of equal-mass systems are degenerate with single sources of the same fundamental parameters. Our model is capable of identifying overluminosity down to $M_G - A_G \sim 16$. 

The predictions from our model are displayed in Table \ref{tab:model_pred} along with the measured color and absolute magnitude. We include a flag in this table which signals if the \textit{Gaia} source is part of a young moving group in BANYAN $\Sigma$, as overluminosity for these stars can be attributed to youth rather than binarity. Another possible cause of overluminosity is an unusually inflated radius that could be caused by strong magnetic fields which create large star spot coverage: indeed, some low-mass stars are known to have radii 10-15\% too large correlated with 5-10\% smaller temperatures \citep{cao_inflation}. In order to surpass our overluminosity cutoff at $\Delta M \sim 0.376$, the corresponding change in radius would be $\sim 18$ \% by the Stefan-Boltzmann law (assuming constant temperature), much larger than what you would expect from magnetic effects. Both youth and magnetic activity are expected to be rare in the local field population, but may still create additional uncertainties in our overluminosity predictions.

We validate our method with the larger Full Sample using sets of known binaries, including astrometric, eclipsing, and RV variable systems and show that we reliably recover the near equal-mass stars (with the largest magnitude offset) as overluminous. However, this validation highlights that our method does not work well for systems with faint companions. 

We then combine our identified binaries with catalogs of known multiple systems from \textit{Gaia} DR3 to create a catalog of ``systems'' within 100 parsecs. Our catalog is generated by finding associations of \textit{Gaia} \textsc{source\_id}s in catalogs of common proper motion systems. We also use the multiplicity flags from \cite{carmenes_multiplicity} to count the number of unresolved companions. Our resulting \textit{Catalog of Systems} contains 347,440 systems of which 126,073 are found to be multiples. Furthermore, of the 21,864 stars identified as overluminous, 9,741 candidate binaries have no previous indication of an unresolved component.

Finally, we use our catalog to provide a lower limit on the stellar multiplicity for the lower main sequence of stars. After removing systems with white dwarfs we find the multiplicity fraction across different bins of mass, from the $M_{K_s}$ relation given by \cite{mann_mass}. Our multiplicity fraction is roughly consistent across equal volume distance bins except in the $M\in[0.5,0.6]~M_{\odot}$ bin. This may be due to systems missing from the \textit{Catalog of Systems} due to resolution, flux contrast, and selection effects. We observe decreasing multiplicity for lower stellar mass, which agrees with the results from other works. However, when compared to volume complete samples we are likely missing faint companions leading to lower predictions of multiplicity.

Despite only providing lower limits to the stellar multiplicity fraction, this work provides a unique method of identifying low-mass multiples in the field population \textit{without} the need for accurate stellar labels. We are able to identify binaries at significantly fainter magnitudes than other methods ($M_G>10$) in the region of the HR diagram where stellar models still struggle. Conversely, since we are able to identify overluminous stars this work provides a set of low-mass stars which are effectively single systems with companions too faint to appear in the photometry. 


\section{Acknowledgements}
ZW would like to acknowledge the contributions to this work that came from conversations with Wei-Chun Jao, Russel White, Viacheslav (Slava) Sadykov, Gregory M. Green, David Hogg, Zachary Hartman, Melodie Sloeneker, Kayvon Sharifi, Todd Henry, Aman Kar, Madison LeBlanc, Tim Johns, and Sebasti\'an Carrazco-Gaxiola.

We acknowledge with thanks the variable star observations from the AAVSO International Database contributed by observers worldwide and used in this research.

This research made use of the Montreal Open Clusters and Associations (MOCA) database, operated at the Montr\'eal Plan\'etarium (J. Gagn\'e et al., in preparation).

This work has made use of data from the European Space Agency (ESA) mission {\it Gaia} (\url{https://www.cosmos.esa.int/gaia}), processed by the {\it Gaia} Data Processing and Analysis Consortium (DPAC, \url{https://www.cosmos.esa.int/web/gaia/dpac/consortium}). Funding for the DPAC has been provided by national institutions, in particular the institutions participating in the {\it Gaia} Multilateral Agreement.

This publication makes use of data products from the Two Micron All Sky Survey, which is a joint project of the University of Massachusetts and the Infrared Processing and Analysis Center/California Institute of Technology, funded by the National Aeronautics and Space Administration and the National Science Foundation.

This publication makes use of data products from the Wide-field Infrared Survey Explorer, which is a joint project of the University of California, Los Angeles, and the Jet Propulsion Laboratory/California Institute of Technology, funded by the National Aeronautics and Space Administration.

Guoshoujing Telescope (the Large Sky Area Multi-Object Fiber Spectroscopic Telescope LAMOST) is a National Major Scientific Project built by the Chinese Academy of Sciences. Funding for the project has been provided by the National Development and Reform Commission. LAMOST is operated and managed by the National Astronomical Observatories, Chinese Academy of Sciences.

\software{
Astropy \citep{astropy:2013, astropy:2018, astropy:2022},
SciPy \citep{2020SciPy-NMeth},
NumPy \citep{harris2020array},
scikit-learn \citep{scikit-learn},
matplotlib \citep{Hunter:2007},
pandas \citep{reback2020pandas, mckinney-proc-scipy-2010},
gaiaxpy (\url{https://gaia-dpci.github.io/GaiaXPy-website/}),
Anaconda Software Distribution (\url{https://www.anaconda.com}),
}


\appendix
\section{Regression Sample Query}
\label{appendix:regression_sample}
The following query was used to generate the 138,600 star sample from the \textit{Gaia} archive. This sample of stars was further cleaned using the color excess $C^*$ in \S\ref{sec:data}.
\begin{verbatim}
SELECT * FROM gaiadr3.gaia_source AS gs
JOIN gaiadr3.xp_summary AS xp 
    USING (source_id)
WHERE gs.g_rp>0.56
AND gs.phot_g_mean_mag + 
    5*log10(gs.parallax/1000) + 
    5 > 5.553
AND gs.phot_g_mean_mag + 
    5*log10(gs.parallax/1000) + 
    5 < 10*gs.g_rp+5
AND gs.parallax > 10
AND gs.parallax_over_error > 10
AND gs.has_xp_continuous = 'True' 
AND CAST(xp.rp_n_contaminated_transits 
        AS FLOAT)/
    CAST(xp.rp_n_transits AS FLOAT) < 0.1
AND CAST(xp.rp_n_blended_transits 
        AS FLOAT)/
    CAST(xp.rp_n_transits AS FLOAT) < 0.1
AND gs.ruwe < 1.2
\end{verbatim}

\section{Full Sample Query}
\label{appendix:full_sample}
The following query was used to generate the 183,884 star sample from the \textit{Gaia} archive. This sample is discussed in \S\ref{sub:train_model}
\begin{verbatim}
SELECT * FROM gaiadr3.gaia_source AS gs
JOIN gaiadr3.xp_summary AS xp 
    USING (source_id)
WHERE gs.bp_rp>0.983
AND gs.g_rp>0.56
AND gs.phot_g_mean_mag + 
    5*log10(gs.parallax/1000) + 
    5 > 5.553
AND gs.phot_g_mean_mag + 
    5*log10(gs.parallax/1000) + 
    5 < 10*gs.g_rp+5
AND gs.parallax > 10
AND gs.parallax_over_error > 10
AND gs.has_xp_continuous = 'True' 
AND CAST(xp.rp_n_contaminated_transits 
        AS FLOAT)/
    CAST(xp.rp_n_transits AS FLOAT) < 0.1
AND CAST(xp.rp_n_blended_transits 
        AS FLOAT)/
    CAST(xp.rp_n_transits AS FLOAT) < 0.1
\end{verbatim}

\bibliography{my_bib}{}
\bibliographystyle{aasjournal}



\end{document}